\documentclass[english,pre,showpacs]{revtex4}
\usepackage[T1]{fontenc}
\usepackage[latin9]{inputenc}
\usepackage{booktabs}
\usepackage{units}
\usepackage{amsmath}
\usepackage{graphicx}
\usepackage{amssymb}
\usepackage{esint}

\makeatletter

\providecommand{\tabularnewline}{\\}
\newcommand{\lyxdot}{.}

\@ifundefined{textcolor}{}
{%
 \definecolor{BLACK}{gray}{0}
 \definecolor{WHITE}{gray}{1}
 \definecolor{RED}{rgb}{1,0,0}
 \definecolor{GREEN}{rgb}{0,1,0}
 \definecolor{BLUE}{rgb}{0,0,1}
 \definecolor{CYAN}{cmyk}{1,0,0,0}
 \definecolor{MAGENTA}{cmyk}{0,1,0,0}
 \definecolor{YELLOW}{cmyk}{0,0,1,0}
 }

\DeclareMathOperator{\im}{Im}

\makeatother

\usepackage{babel}

\begin{document}

\title{Quantum chaos of a mixed, open system of kicked cold atoms}

\author{Yevgeny Krivolapov}

\affiliation{Physics Department, Technion - Israel Institute of Technology, Haifa
32000, Israel.}

\email{evgkr@tx.technion.ac.il}

\author{Shmuel Fishman}

\affiliation{Physics Department, Technion - Israel Institute of Technology, Haifa
32000, Israel.}

\author{Edward Ott}

\affiliation{University of Maryland, College Park, Maryland 20742, USA}

\author{Thomas M. Antonsen}

\affiliation{University of Maryland, College Park, Maryland 20742, USA}


\begin{abstract}
The quantum and classical dynamics of particles kicked by a gaussian
attractive potential are studied. Classically, it is an open mixed
system (the motion in some parts of the phase space is chaotic, and
in some parts it is regular). The fidelity (Loschmidt echo) is found
to exhibit oscillations that can be determined from classical considerations
but are sensitive to phase space structures that are smaller than
Planck's constant. Families of quasi-energies are determined from
classical phase space structures. Substantial differences between
the classical and quantum dynamics are found for time dependent scattering.
It is argued that the system can be experimentally realized by cold
atoms kicked by a gaussian light beam.
\end{abstract}
\maketitle

\section{Introduction}

The quantum behavior of classically chaotic systems has been extensively
studied both with time dependent and time independent Hamiltonians
\cite{Tabor1989,OzoriodeAlmeida1988,Varenna1991,Gutzwiller1990,Oppo1994,Haake2001,Ott2002}.
The main issue is that of determining fingerprints of classical chaos
in the quantum mechanical behavior. For example, the spectral statistics
of closed classically integrable \cite{Berry1976,Berry1977a,Berry1977b}
and classically chaotic \cite{Bohigas1984,Sieber2001,MullerHaake2004}
quantum systems have been predicted to have clearly distinct properties.
Many of the systems that are of physical interest are mixed, where
some parts of the phase space are chaotic and some parts are regular.
Spectral properties of mixed systems with time independent Hamiltonians
were studied by Berry and Robnik \cite{Berry1984}. In the present
paper we study the classical/quantum correspondence properties of
a mixed, open, time dependent system. (Here by {}``open'' we mean
that both position and momentum are unbounded.)

The system we study consists of a particle kicked by a Gaussian potential
defined by the Hamiltonian,\begin{equation}
H=\frac{p^{2}}{2m}-K'Te^{-\frac{x^{2}}{2\Delta^{2}}}\sum_{n=-\infty}^{\infty}\delta\left(t-Tn\right).\label{eq:Hamiltonian_with_units}\end{equation}
Models of this form were studied by Jensen who used it to investigate
quantum effects on scattering in classically chaotic \cite{Jensen1994}
and mixed \cite{Jensen1992} systems. This system can be experimentally
approximated by a Gaussian laser beam acting on a cloud of cold atoms,
somewhat similar to the realization of the kicked rotor by Raizen
and coworkers \cite{MooreRaizen1995}. As we will show, the study
of Hamiltonian \eqref{eq:Hamiltonian_with_units} is particularly
suited to the investigation of generic behavior of kicked, open, mixed-phase-space
systems. In particular, we will focus on issues of fidelity \cite{Peres1984,Jalabert2001,Jacquod2002,Gorin2006},
decoherence \cite{HansonOtt1984a,Cohen1991} and scattering \cite{Blummel1988,Jalabert1990,Doron1991,Lai1992}.
Our main motivation in studying Hamiltonian \eqref{eq:Hamiltonian_with_units}
is that, with likely future technological advances (see Sec. \ref{sec:Discussion}
for discussion), the phenomena we consider may soon become accessible
to experimental investigation.

Quantum mechanically, it is expected that classical phase space details
on the scale of Planck's constant are washed out \cite{Berry1972,Berry1977c}.
In contrast, one of our results will be that quantum dynamics can
be sensitive to extremely fine structures in phase space, and this
sensitivity is stable in the presence of noise \cite{HansonOtt1984a,Cohen1991}.
Phase space tunneling has been studied extensively \cite{Hanson1984,Tomsovic1994,Sheinman2006,Lock2010}.
For systems with many phase space structures complications arise due
to transport between these structures. For our Hamiltonian \eqref{eq:Hamiltonian_with_units}
the motion is unbounded (i.e., the system is {}``open''), and therefore
this system is ideal for the exploration of tunneling out of phase
space structures and, in particular, for study of resonance assisted
tunneling, a current active field of research \cite{Tomsovic1994,Sheinman2006,Lock2010}.

The outline of our paper is as follows. Section \ref{sec:The-model}
presents and discusses our model system. Section \ref{sec:Quasi-energies-of-an}
considers the quasi-energies of quantum states localized to island
chains. Section \ref{sec:Fidelity} introduces the fidelity concept
and applies it to study different regions of the phase space including
the main, central KAM island (Sec. \ref{sub:Fidelity_central}), island
chains (Sec. \ref{sub:Fidelity_chain}), and chaotic regions (Sec.
\ref{sub:Fidelity_chaos}). Experimentally there is always some noise
present in such systems. Also, noise can be intentionally introduced.
Section \ref{sec:Dephasing} considers this issue. Section \ref{sec:Scattering}
presents a study of the scattering properties of the system. Conclusions
and further discussion are given in Sec. \ref{sec:Discussion}.

\section{\label{sec:The-model}The model}

\subsection{Formulation}

A particle kicked by a Gaussian beam is modeled by the Hamiltonian,
Eq. \eqref{eq:Hamiltonian_with_units}, with the classical equations
of motion,\begin{eqnarray}
\dot{p} & = & -\frac{\partial H}{\partial x}=-K'T\frac{x}{\Delta^{2}}e^{-\frac{x^{2}}{2\Delta^{2}}}\sum_{n=-\infty}^{\infty}\delta\left(t-Tn\right),\label{eq:Ham_eq_motion}\\
\dot{x} & = & \frac{\partial H}{\partial p}=\frac{p}{m}.\nonumber \end{eqnarray}
We rewrite the equations of motion in dimensionless form by defining
$\bar{x}=x/\Delta$, $\bar{t}=t/T$. The dimensionless momentum is
correspondingly defined as $\bar{p}=pT/\left(m\Delta\right)$. Thus
we obtain the dimensionless equations of motion,\begin{eqnarray}
\dot{\bar{p}} & = & -K\bar{x}e^{-\frac{\bar{x}^{2}}{2}}\sum_{n=-\infty}^{\infty}\delta\left(\bar{t}-n\right),\\
\dot{\bar{x}} & = & \bar{p},\nonumber \end{eqnarray}
where\begin{equation}
K=\frac{K'T^{2}}{m\Delta^{2}}.\end{equation}
Since in what follows we deal with the rescaled position, momentum
and time we will drop the bar notation for convenience. By integrating
\eqref{eq:Ham_eq_motion} and defining $p_{n}=p\left(t=n_{-}\right)$,
$x_{n}=x\left(t=n_{-}\right)$, where $t=n_{-}$ is a time just before
the \emph{n-}th kick, we can rewrite the differential equations of
the motion as a mapping, $M$,\begin{equation}
M:\qquad\begin{cases}
p_{n+1} & =p_{n}-Kx_{n}e^{-\frac{x_{n}^{2}}{2}},\\
x_{n+1} & =x_{n}+p_{n+1}.\end{cases}\label{eq:classical_map}\end{equation}
The corresponding quantum dynamics in rescaled units is given by the
Hamiltonian,\begin{equation}
H=\frac{p^{2}}{2}-Ke^{-\frac{x^{2}}{2}}\sum_{n=-\infty}^{\infty}\delta\left(t-n\right),\label{eq:Hamiltonian}\end{equation}
where $p=-i\tau\partial_{x}$, and $\tau=\hbar T/\left(m\Delta^{2}\right)$
is the rescaled $\hbar$, namely, $\left[x,p\right]=i\tau$. The quantum
evolution is given by\begin{equation}
i\tau\partial_{t}\psi=-\frac{\tau^{2}}{2}\partial_{xx}\psi-Ke^{-\frac{x^{2}}{2}}\sum_{n=-\infty}^{\infty}\delta\left(t-n\right)\psi,\end{equation}
or by the one kick propagator\begin{equation}
U_{1}=e^{-i\frac{p^{2}}{2\tau}}\exp\left(i\frac{K}{\tau}e^{-\frac{x^{2}}{2}}\right).\end{equation}

\subsection{Properties of the Classical Map and the phase portrait}

\begin{figure}
\centering{}\includegraphics[width=8cm]{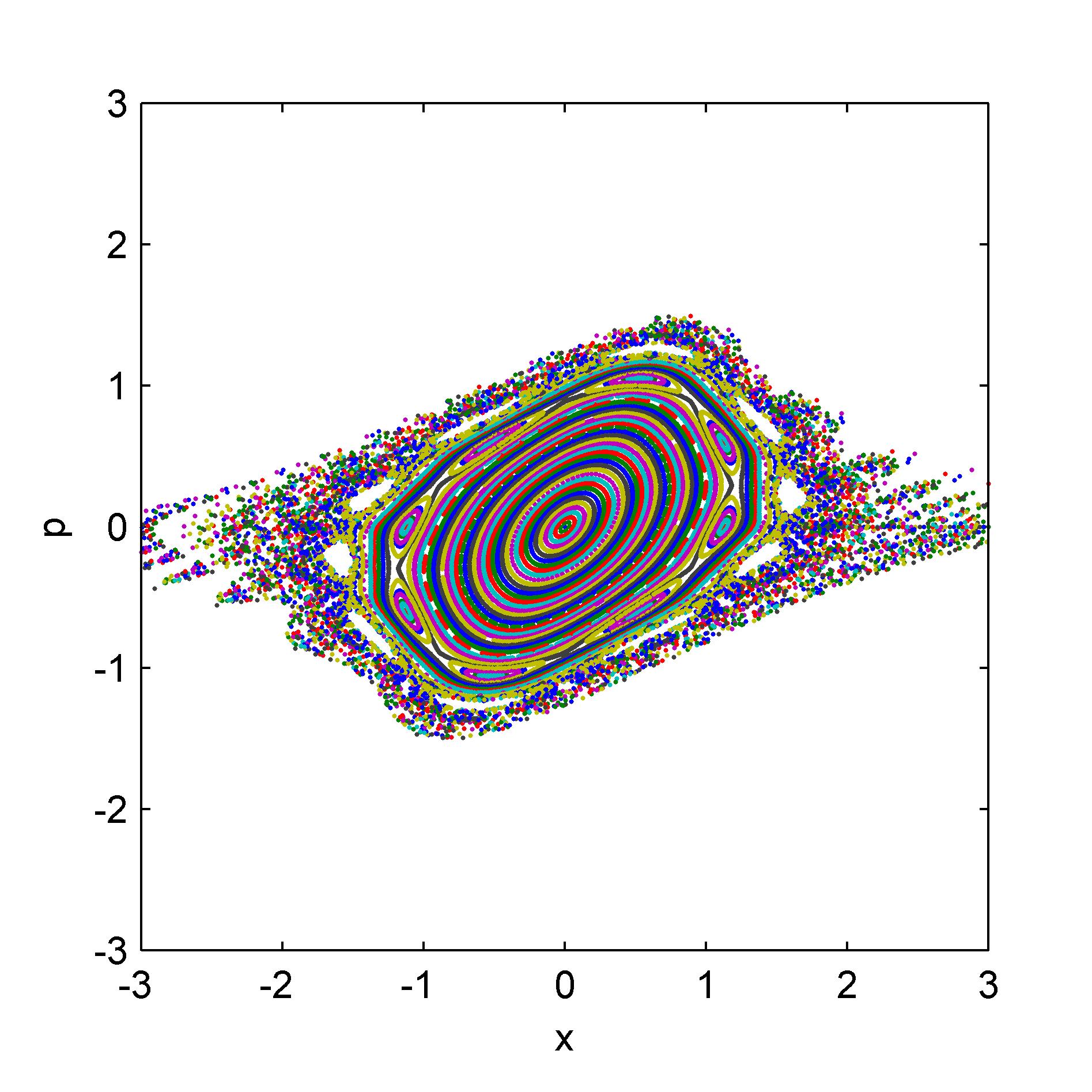}\caption{\label{fig:phase_space_K1}(Color online) The phase space for $K=1$.
Colors (shades) distinguish different orbits.}

\end{figure}

\begin{figure}
\centering{}\includegraphics[width=8cm]{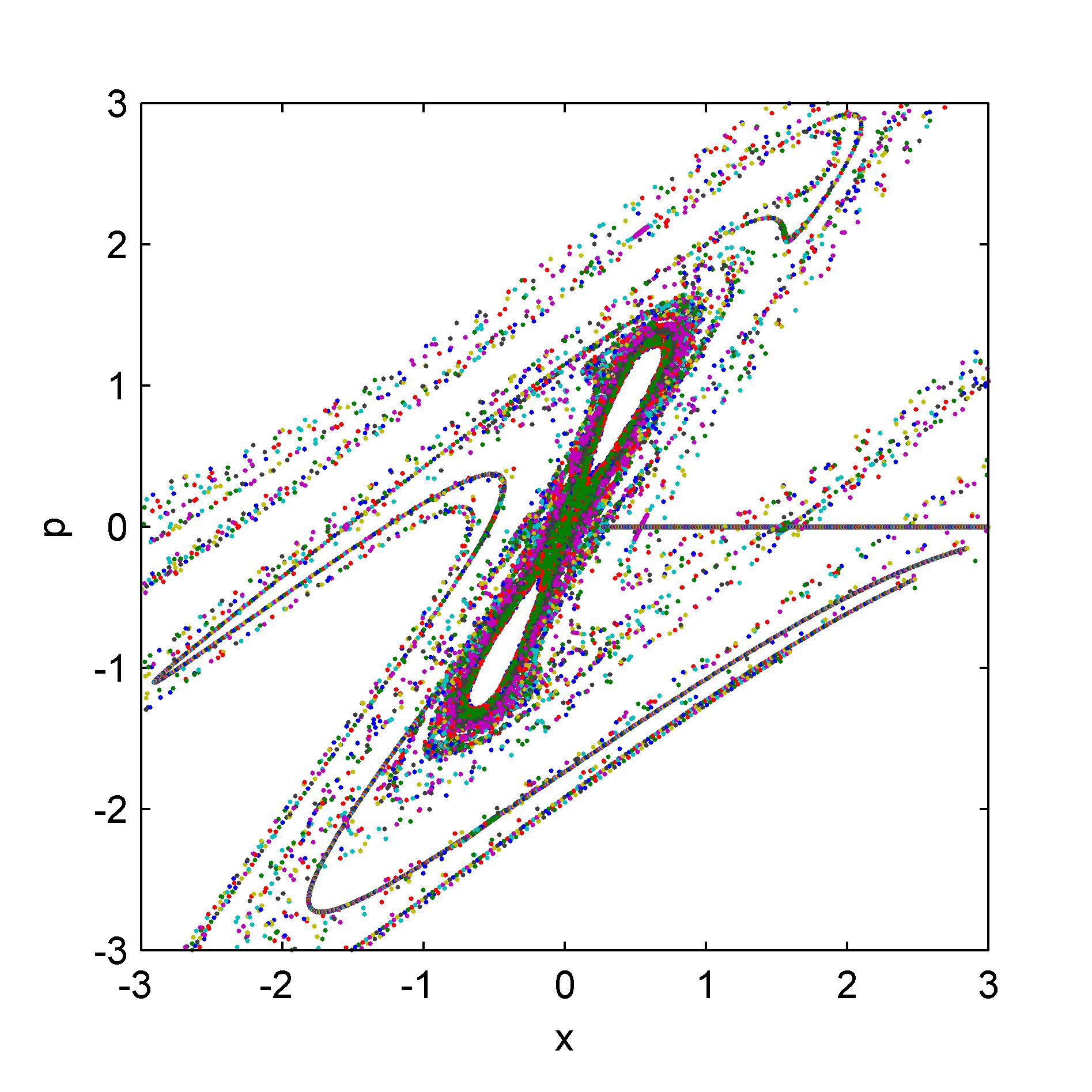}\caption{\label{fig:phase_space_K4.5}(Color online) The phase space for $K=4.5$.
Colors (shades) distinguish different orbits.}

\end{figure}

In this subsection the classical properties of the map Eq. \eqref{eq:classical_map}
will be presented. The first property is reflection symmetry, $\left(x,p\right)\rightarrow\left(-x,-p\right)$.
Phase portraits such as these presented in Fig. \ref{fig:phase_space_K1}
and Fig. \ref{fig:phase_space_K4.5} %
are clearly seen to satisfy this property. Like the standard map,
Eq. \eqref{eq:classical_map} can be written as a product of two involutions,
$M=J_{2}J_{1}$, where\begin{eqnarray}
J_{1}:\qquad\left(p,x\right) & \rightarrow & \left(-p,x+p\right)\label{eq:involutions}\\
J_{2}:\qquad\left(p,x\right) & \rightarrow & \left(-p-Kxe^{-\frac{x^{2}}{2}},x\right).\nonumber \end{eqnarray}
We will use this property for the calculation of the periodic orbits.
From \eqref{eq:classical_map} we see that the only fixed point is
$\left(x=0,p=0\right)$. Linearizing around this point, we find that
the trace of the tangent map is $2-K$. Therefore, this point is elliptic
for $0<K<4$, and, for $K=1$, the phase portrait of Fig. \ref{fig:phase_space_K1}
is found, while for $K>4$ this point is hyperbolic, leading to phase
portraits like that of Fig. \ref{fig:phase_space_K4.5}. Since the
kicking as a function of $x$ is bounded by $K$, for large initial
momentum the particle is nearly not affected by the kicks, and continues
to move in its initial direction. For $0<K<4$ we find a large island
around the elliptic point $\left(x,p\right)=\mbox{\ensuremath{\left(0,0\right)}}$,
and, for nearly all initial conditions near $x=p=0$, the motion is
regular (i.e., lies on KAM surfaces). Further away from this fixed
point, one finds island chains embedded in a chaotic strip. And even
further away, the motion is unbounded.

\section{\label{sec:Quasi-energies-of-an}Quasi-energies of an island chain}

In the semiclassical regime quasi-energies are related to classical
structures. In this section we assume the existence of quasi-energy
eigenfunctions $u_{n}\left(x\right)$,\begin{equation}
U_{1}u_{n}\left(x\right)=e^{-iE_{n}}u_{n}\left(x\right),\end{equation}
such that $u_{n}\left(x\right)$ is strongly localized to an island
chain of order $r$, and we attempt to calculate the quasi-energy
$E_{n}$. For this purpose we use the one-kick propagator $U_{1}$
to generate successive jumps in the island chain,\begin{equation}
U_{1}\psi_{i}=\psi_{i+1},\end{equation}
where $\psi_{i}$ is a wavefunction which is localized in island number
$i$ within the island chain. Further, we assume that this wavefunction
can be expanded using the quasi-eigenstates of the island chain,\begin{equation}
\psi_{i}=\sum_{n}c_{in}u_{n}\left(x\right).\label{eq:initial_wavepacket}\end{equation}
Using this expansion we obtain a system of equations,\begin{equation}
U_{1}\psi_{i}=\sum_{n}c_{in}u_{n}\left(x\right)e^{-iE_{n}},\end{equation}
and\begin{equation}
U_{1}^{r}\psi_{i}=\sum_{n}c_{in}u_{n}\left(x\right)e^{-iE_{n}r}.\end{equation}
Classically the $i-$th island is transformed to itself by $r$ successive
applications of the map $M$. In particular, an elliptic fixed point
of the map $M^{r}$ is located in the center of the island. In the
semiclassical limit the eigenstates of $U_{1}^{r}$ are determined
by $M^{r}$ and are close to the eigenstates of a harmonic oscillator
centered on the fixed point of $M^{r}$. The frequency of the oscillator,
$\nu_{i}$, is such that the eigenvalues of the tangent map of $M^{r}$,
which transforms the $i-$th island to itself, are $e^{\pm i\nu_{i}}$.
This tangent map can be written in terms of the product of the tangent
maps of $M\left(i\rightarrow i+1\right),$ which transform the $i-$th
island to the $\left(i+1\right)-$th island. Consequently, since the
eigenvalues are determined by the trace of the product of the tangent
maps, they are independent of $i$ (due to the invariance of the trace
to cyclic permutations). In what follows we therefore drop the index
$i$ from $\nu_{i}$.

Choosing $\psi_{i}$ as the eigenstate of $U_{1}^{r}$, means that\begin{equation}
U_{1}^{r}\psi_{i}=e^{i\bar{\beta}}\psi_{i},\end{equation}
where $\bar{\beta}=\tau\nu/2$ and we have taken $\psi_{i}$ to be
the ground state of the harmonic oscillator. Therefore,\begin{equation}
\psi_{i}=\sum_{n}c_{in}u_{n}\left(x\right)e^{-i\left(E_{n}r+\bar{\beta}\right)}.\label{eq:16}\end{equation}
Using the orthogonality of the $u_{n}\left(x\right)$, Eqs. \eqref{eq:16}
and \eqref{eq:initial_wavepacket} yield\begin{equation}
e^{-i\left(E_{n}r+\bar{\beta}\right)}=1.\label{eq:quasien_cond}\end{equation}
The quasi-energies, obtained from \eqref{eq:quasien_cond} are, therefore,\begin{equation}
E_{n}=\frac{2\pi}{r}n+\beta,\qquad0\leq n\leq r,\label{eq:quasi_en_formula}\end{equation}
where $\beta=-\bar{\beta}/r$. Approximations to the quasi-energies
can be calculated numerically by launching a wavepacket into one island
in the island chain and propagating it in time, which gives\begin{equation}
\psi\left(x,N\right)=\sum_{n}c_{in}u_{n}\left(x\right)e^{-iE_{n}N}.\end{equation}
Taking a Fourier transform with respect to $N$ gives the quasi-energies.
We have found that for $K=1$ the chains with $r=8$ and $r=16$ accurately
satisfy \eqref{eq:quasi_en_formula}.

\section{\label{sec:Fidelity}Fidelity}

The concept of quantum fidelity was introduced by Peres \cite{Peres1984}
as a fingerprint of classical chaos in quantum dynamics. It has subsequently
been extensively utilized in theoretical \cite{Jalabert2001,Jacquod2002,Cerruti2002,Wimberger2006}
and experimental studies \cite{Andersen2004,Kaplan2005,Wimberger2006,Andersen2006},
for a review see \cite{Gorin2006}. Most of this research has focused
on the difference between chaotic and regular systems. Here we discuss
fidelity for a mixed system. We have calculated the fidelity,\begin{equation}
S\left(t\right)=\left|\left\langle \phi_{0}\right|e^{iH_{1}t/\tau}e^{-iH_{2}t/\tau}\left|\phi_{0}\right\rangle \right|^{2},\label{eq:Fidelity}\end{equation}
where $H_{1,2}$ are Hamiltonians of the form \eqref{eq:Hamiltonian}
with with slightly different kicking strengths, $K_{1,2}$, and $\phi_{0}$
is the initial wavefunction. We note that the fidelity $S\left(t\right)$
can be experimentally measured by the Ramsey method, as used in Ref.
\cite{Kaplan2005}. The fidelity is related to an integral over Wigner
functions,\begin{equation}
S\left(t\right)=2\pi\tau\int dxdp\, P_{\phi_{1}}\left(x,p\right)P_{\phi_{2}}\left(x,p\right),\label{eq:Fidelity_Wigner}\end{equation}
where $P_{\phi_{1,2}}$ are the Wigner functions of $\phi_{1,2}=e^{-iH_{1,2}t/\tau}\phi_{0}$,
respectively.

We study separately the fidelity in the central island, in the island
chain, and in the chaotic region (i.e., Eq.\eqref{eq:Fidelity} with
$\phi_{0}$ localized to these regions).

\subsection{\label{sub:Fidelity_central}Fidelity of a wavepacket in the central
island}

\begin{figure}
\begin{centering}
\includegraphics[width=8cm]{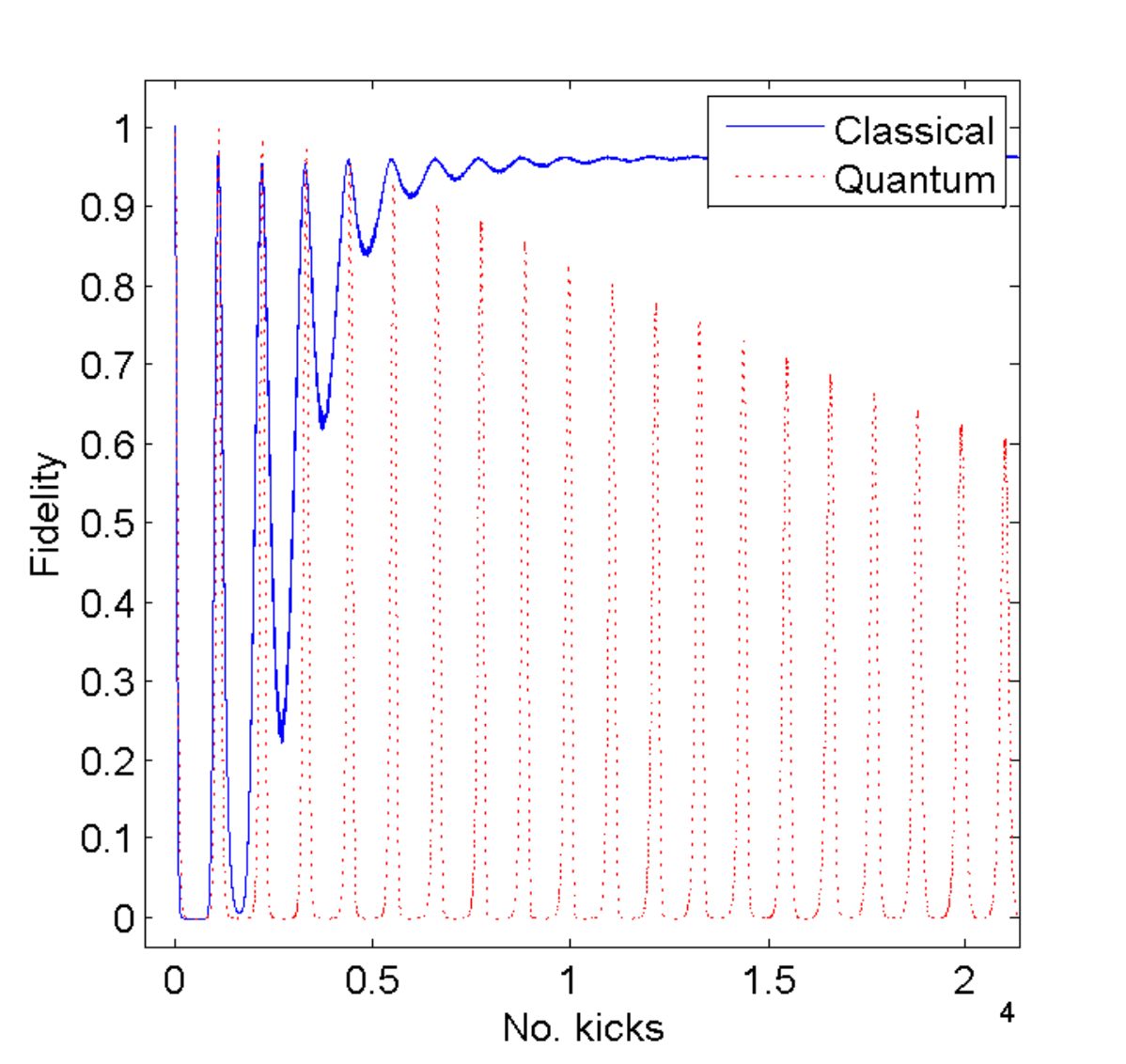}
\par\end{centering}

\caption{\label{fig:Quantum_classical_fidelity}(Color online) Quantum fidelity,
$S\left(t\right)$, (dashed red) and classical fidelity, $S_{c}\left(t\right)$
(solid blue). $K_{1}=1$, $K_{2}=1.01$, $\tau=0.01$, $x_{0}=-0.25$
and $p_{0}=0$.}

\end{figure}

\begin{figure}
\begin{centering}
\includegraphics[width=8cm]{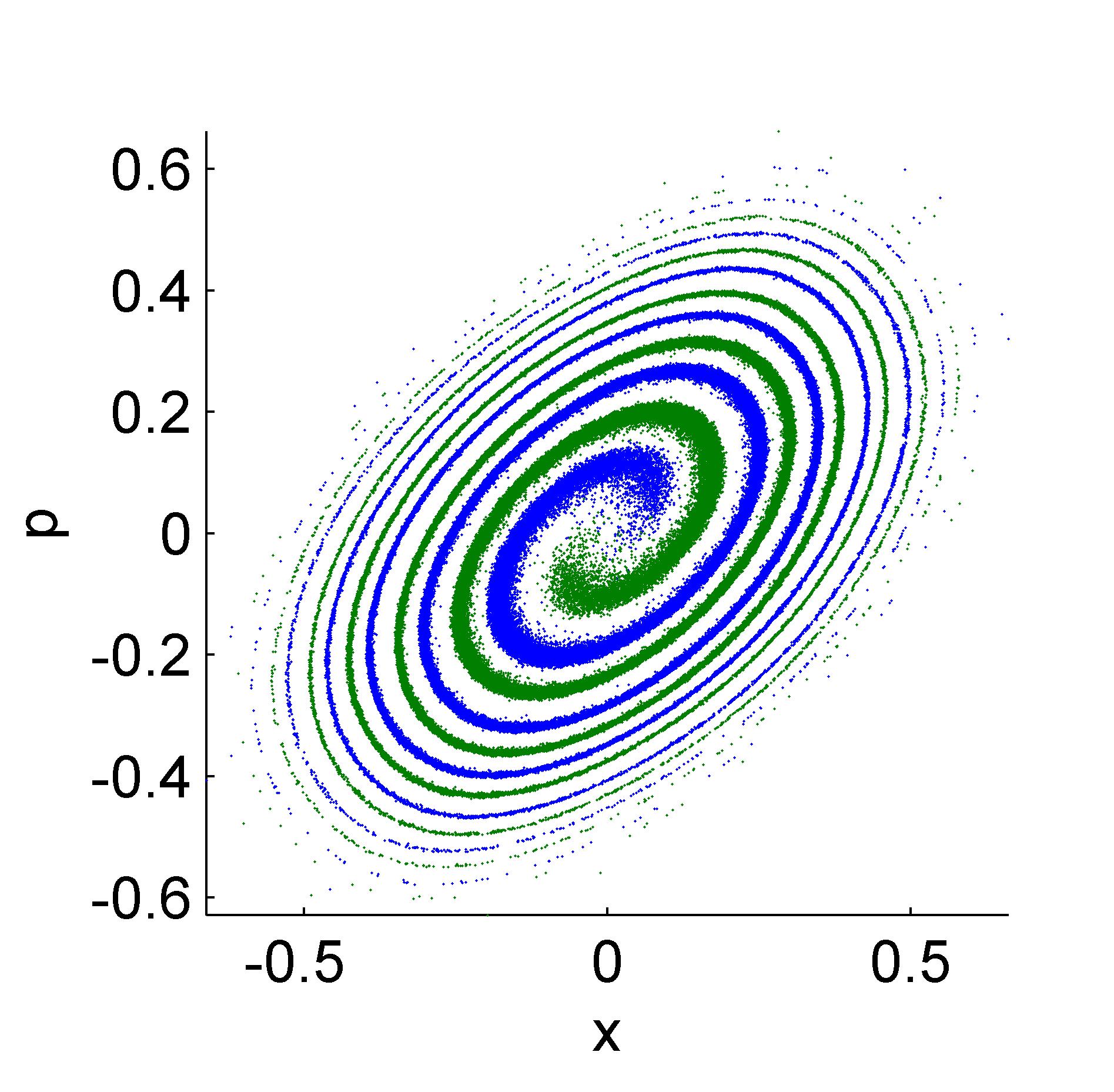}
\par\end{centering}

\caption{\label{fig:Whorls}(Color online) Classical density, which was initially
placed at $x_{0}=-0.25$ and $p_{0}=0$ after $500$ kicks. Blue (dark)
dots are for $K_{1}=1$ and green (light) dots are for $K_{2}=1.01$.}

\end{figure}

\begin{figure}
\begin{centering}
\includegraphics[width=8cm]{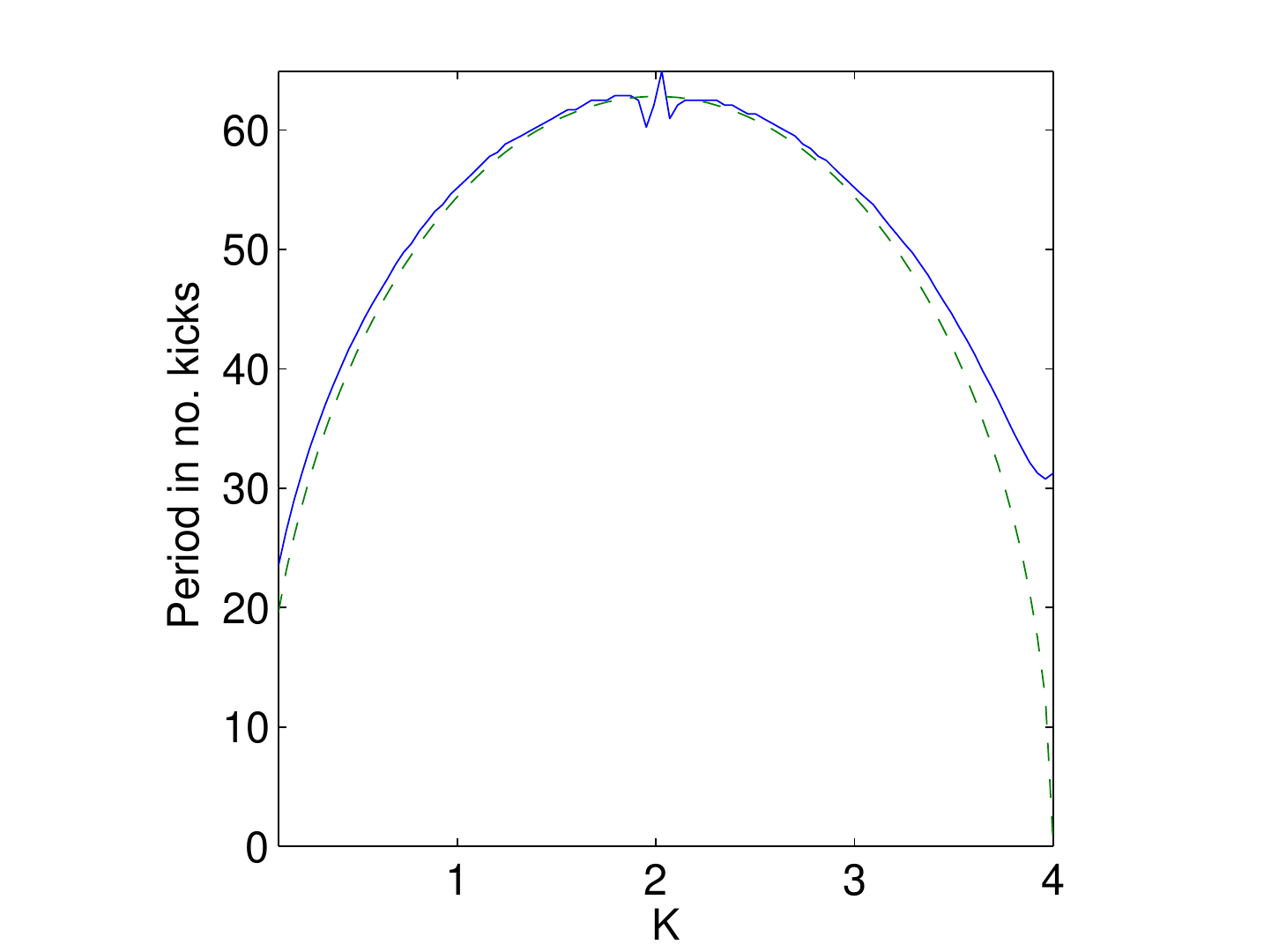}
\par\end{centering}

\caption{\label{fig:Fidelity_center}(Color online) A numerical \.{(}solid
blue) and an analytical (dashed green) computation of the period of
the fidelity revival as a function of $K$, $\delta K=0.1$, $x_{0}=p_{0}=0$,
$\tau=0.01$.}

\end{figure}

First we prepare the initial wavefunction $\phi_{0}$ as a Gaussian
wavepacket with a minimal uncertainty, namely, $\Delta x=\Delta p=\left(\tau/2\right)^{\nicefrac{1}{2}}$,
\begin{equation}
\phi_{0}\left(x\right)=\frac{1}{\left(2\pi\left(\Delta x\right)^{2}\right)^{\nicefrac{1}{4}}}e^{-ip_{0}x/\tau}\exp\left[-\frac{\left(x-x_{0}\right)^{2}}{4\left(\Delta x\right)^{2}}\right].\label{eq:Gaussian_quantum}\end{equation}
We place $\phi_{0}$ in the center of the island, namely, $x_{0}=p_{0}=0$.
Since the center of the wavepacket is initially at the fixed point,
for $\Delta x$ and $\Delta p$ classically small, its dynamics are
approximately determined by the tangent map of the fixed point. For
this purpose we linearize the classical map \eqref{eq:classical_map}
around the point $x=p=0$. This gives the equation for the deviations,\begin{equation}
\left(\begin{array}{c}
\delta x_{n+1}\\
\delta p_{n+1}\end{array}\right)=\left(\begin{array}{cc}
\left(1-K\right) & 1\\
-K & 1\end{array}\right)\left(\begin{array}{c}
\delta x_{n}\\
\delta p_{n}\end{array}\right).\end{equation}
The eigenvalues of this equation are,\begin{equation}
\alpha_{1,2}=\frac{\left(2-K\right)}{2}\pm\frac{i\sqrt{K\left(4-K\right)}}{2}\equiv e^{\pm i\omega},\end{equation}
with\begin{equation}
\omega=\arctan\frac{\sqrt{K\left(4-K\right)}}{\left(2-K\right)},\end{equation}
which is the angular velocity of the points around the origin. In
the vicinity of the fixed point, the system behaves like a harmonic
oscillator with a frequency $\omega$. Classically, the motion of
the trajectories, starting near the elliptic fixed point, $x=p=0$,
stays there because the region is bounded by KAM curves that surround
this point. For small effective Planck's constant, $\tau$, the quantum
behavior is expected to mimic the classical behavior for a long time.
Inspired by the relation between the fidelity and the Wigner function
(see \eqref{eq:Fidelity_Wigner}), we have defined a classical fidelity,
$S_{c}\left(t\right)$, as the overlap between coarse-grained Liouville
densities of $H_{1}$ and $H_{2}$ (this is similar to the classical
fidelity defined in \cite{Nielsen2000}). To do this we first randomly
generate a large number of initial classical positions using the initial
distribution function,\begin{equation}
f_{0}\left(x,p\right)=\frac{1}{2\pi\Delta x\Delta p}\exp\left\{ -\frac{1}{2}\left[\left(\frac{x-x_{0}}{\Delta x}\right)^{2}+\left(\frac{p-p_{0}}{\Delta p}\right)^{2}\right]\right\} ,\label{eq:Gaussian_classical}\end{equation}
corresponding to our initial $\phi_{0}$ given by \eqref{eq:Gaussian_quantum}.
The coarse grained densities for $H_{1}$ and $H_{2}$ are then computed
by first integrating these initial conditions and then coarse graining
to a grid of squares in phase space of area $\tau$ \cite{Berry1979}.
The motivation for this procedure is to check if structures in phase
space of size smaller than $\tau$ are of importance to the fidelity.
A comparison between $S\left(t\right)$ and $S_{c}\left(t\right)$
for $x_{0}=-0.25$, $p_{0}=0$, and $\tau=0.01$ is presented in Fig.
\ref{fig:Quantum_classical_fidelity}.%
 The initial wavepacket is smeared on a ring in the phase space due
to the twist property of the map. Since the probability density is
preserved, the {}``whorl'' which is formed contains very dense and
thin tendrils. In Fig. \ref{fig:Whorls} %
two such {}``whorls'' are presented for $H_{1}$ with $K_{1}=1$
and $H_{2}$ with $K_{2}=1.01$. When the two {}``whorls'' coincide
a fidelity revival is formed. Coarse graining the densities to a boxes
of size $\tau$ averages the differences between the two {}``whorls'',
obtained by $H_{1}$ and $H_{2}$. This explains why the classical
fidelity approaches $1$ as the number of kicks becomes large. On
the other hand, the quantum fidelity shows strong revivals which suggests
that it feels the difference in trajectories between the two Hamiltonians.
To understand the period of the revivals, we calculate, $\delta\omega$,
the frequency difference between the two Hamiltonians, $H_{1,2}$.
Expanding $\omega$ around $K_{1}$ gives\begin{equation}
\omega\left(K\right)=\omega\left(K_{1}\right)+\frac{K-K_{1}}{\sqrt{K\left(4-K\right)}}+O\left(\left(K-K_{1}\right)^{2}\right).\end{equation}
Therefore, the difference in angular velocity between two orbits of
Hamiltonians, $H_{1}$ and $H_{2}$ is given by\begin{equation}
\delta\omega=\omega\left(K_{2}\right)-\omega\left(K_{1}\right)=\frac{\delta K}{\sqrt{K_{2}\left(4-K_{2}\right)}},\end{equation}
for $\delta K=K_{2}-K_{1}$. This suggests that the fidelity, $S\left(t\right)$
will be periodic, with the period $T=\pi/\delta\omega$. Note that
we predict $T=\pi/\delta\omega$, rather than $T=2\pi/\delta\omega$.
This is because of the symmetry of the initial condition. Each point
of $H_{1}$ is chasing a point of $H_{2}$ which is its reflection
through the origin of the phase space and, therefore, is found first
at an angle of $\pi$ and not $2\pi$. To check this, we have calculated
the period of the revivals numerically for $0<K<4$. First, fidelity
was computed and Fourier transformed, then the second most significant
value was taken as the period. In Fig. \ref{fig:Fidelity_center}
we present a comparison of the analytic calculation of the period
of the fidelity and the numerical computation. The correspondence
is good through the whole range of the stochasticity parameter $K$
but degrades near $K=4$, where the elliptic point at the origin becomes
unstable. Also, near $K=2$, resonance chains appear near the fixed
point $x=p=0$, which results in poor agreement with the theoretical
prediction, see Fig. \ref{fig:phase_space_K2.1}.%
 Very often it is assumed that quantum mechanical behavior is insensitive
to phase space structures with areas smaller than Planck's constant,
which results in an effective averaging on this scale \cite{Berry1972,Berry1977c}.
While this assumption is often correct \cite{Sheinman2006}, sometimes
it is not \cite{HensingerPhillips2001,SteckRaizen2001,SteckRaizen2002,AverbukhMoiseyev1995,AverbukhMoiseyev2002,OsovskiMoiseyev2005}.
The difference between $S\left(t\right)$ and $S_{c}\left(t\right)$
demonstrated in Fig. \ref{fig:Quantum_classical_fidelity} shows that
fidelity may be sensitive to extremely small details in the classical
phase space. In particular, a {}``whorl'' \cite{Berry1972,Berry1977c}
affects the quantum dynamics. The small decay of the quantum fidelity
seen in Fig. \ref{fig:Quantum_classical_fidelity} is a result of
tunneling.

We stress that to observe the oscillations which appear on Fig. \ref{fig:Quantum_classical_fidelity}
requires sensitivity to the structure of the {}``whorl'' of Fig.
\ref{fig:Whorls}. In our quantum calculation the effective Planck's
constant is $\tau=0.01$ and it is obvious that the {}``whorl''
of Fig. \ref{fig:Whorls} exhibits structures on smaller scale, for
example, in a square with sides of length $0.1$ in phase space (of
Fig. \ref{fig:Whorls}) one finds several stripes of the {}``whorl''.
Indeed, averaging over such a square leads to the classical fidelity
that does not exhibit oscillations as the quantum fidelity does. We
conclude that the structures on the scale smaller than the effective
Planck's constant, $\tau$, are crucial for the oscillations in the
quantum fidelity. Hence, structures of scales smaller than Planck's
constant may dominate fidelity, which is a quantum quantity.

\begin{figure}
\begin{centering}
\includegraphics[width=8cm]{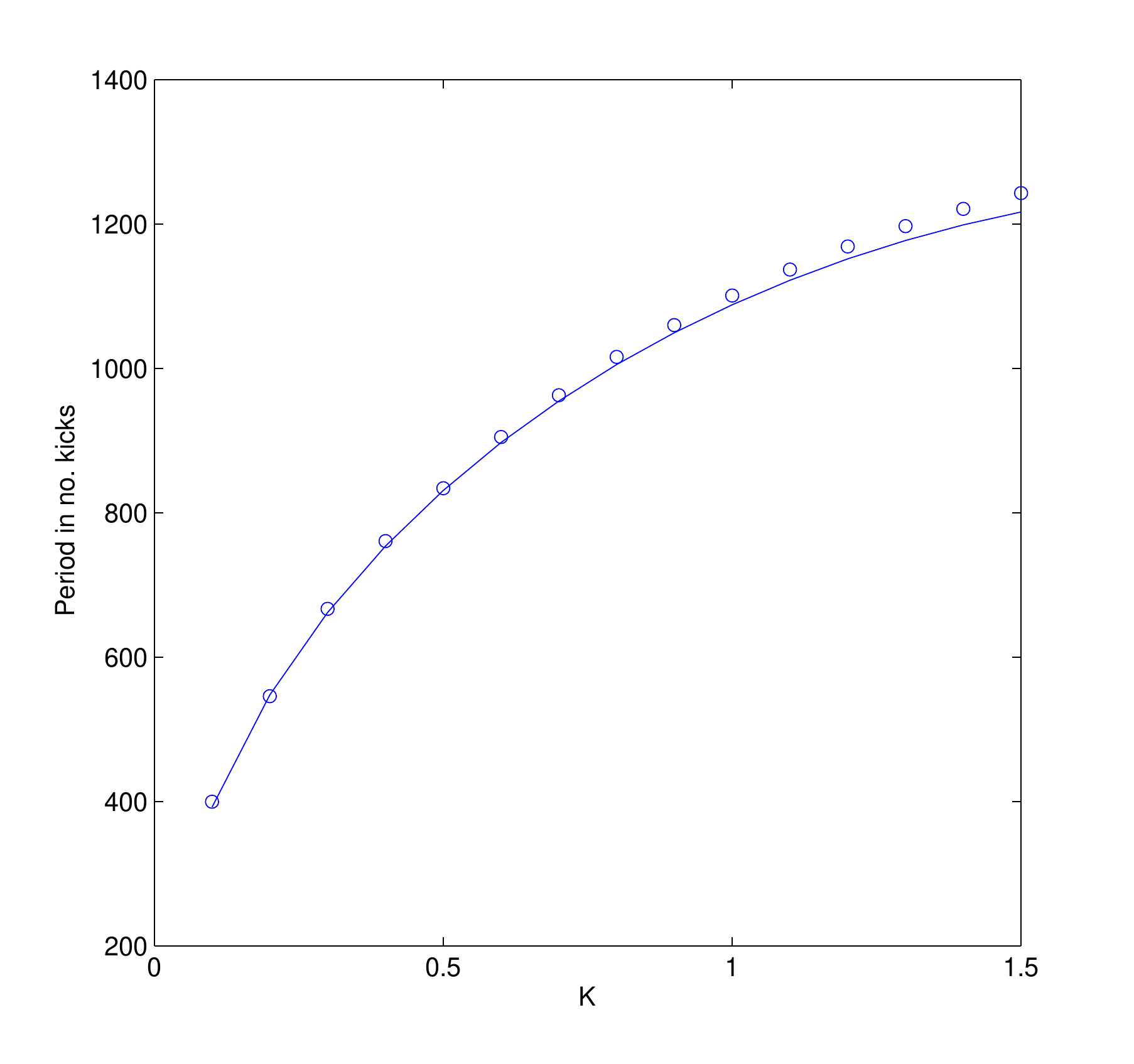}
\par\end{centering}

\caption{\label{fig:Fidelity_near_center}A numerical (blue circles) and an
analytical (solid blue line) computation of the period of the fidelity
revival as a function of $K$, $\delta K=0.01$, $x_{0}=-0.25$, $p_{0}=0$,
$\tau=2\times10^{-4}$.}

\end{figure}

For a wavepacket started around an initial point $\left(x_{0},p_{0}\right)\neq\left(0,0\right)$
the behavior is similar, but with a slightly different period due
to a decrease in the angular velocity for points far from the fixed
point. Similarly to the case of $\left(x_{0},p_{0}\right)=\left(0,0\right)$,
we have calculated numerically the revival period for different values
of $K$; this is shown in Fig. \ref{fig:Fidelity_near_center}%
. For $K>1.5$ resonances appear near the launching point which introduce
additional periods into the fidelity, making the analysis more complicated.

\subsection{\label{sub:Fidelity_chain}Fidelity for a wavepacket in an island
chain}

\begin{figure}
\centering{}\includegraphics[width=8cm]{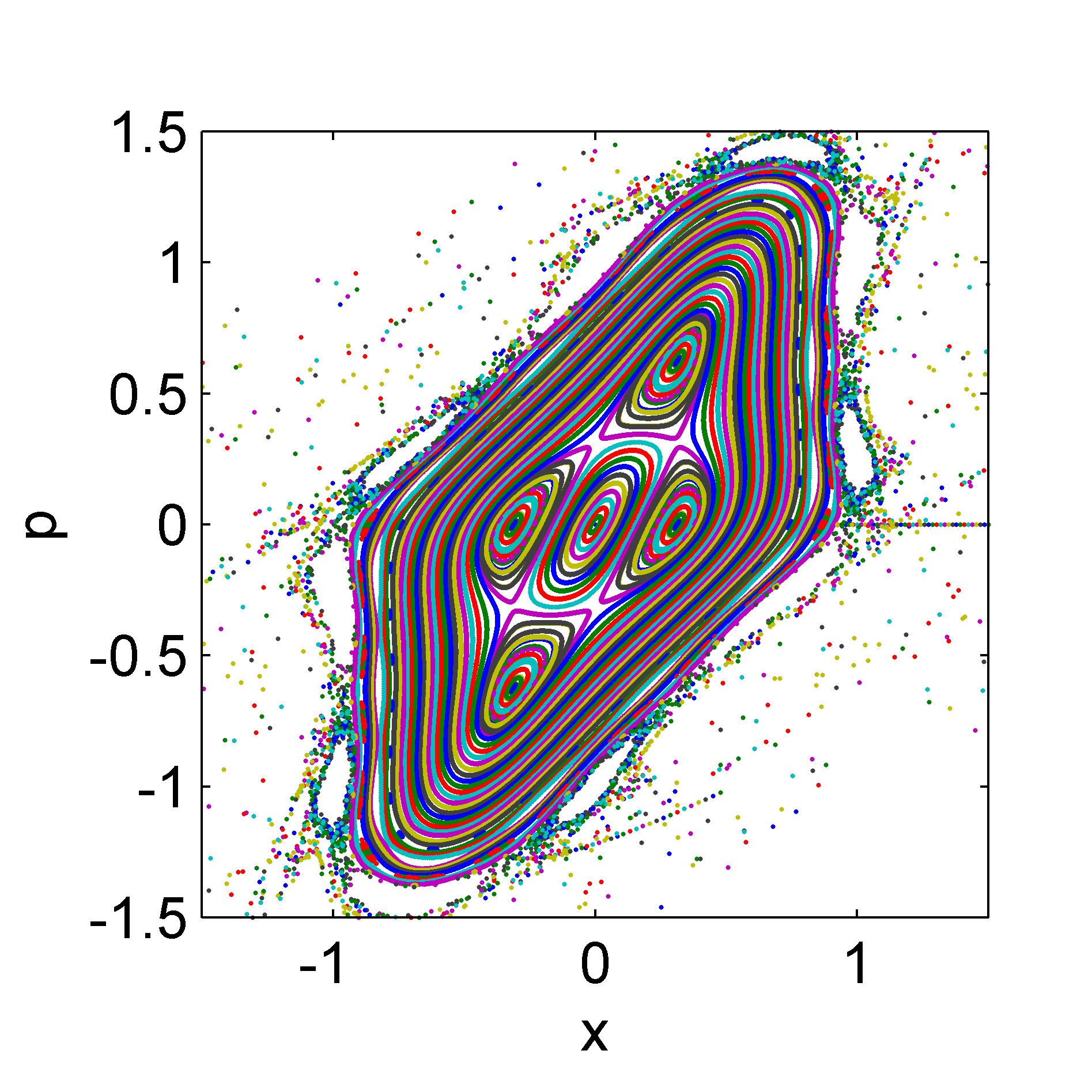}\caption{\label{fig:phase_space_K2.1}(Color online) The phase space for $K=2.1$.
Colors (shades) distinguish different orbits.}

\end{figure}

\begin{figure}
\begin{centering}
\includegraphics[width=8cm]{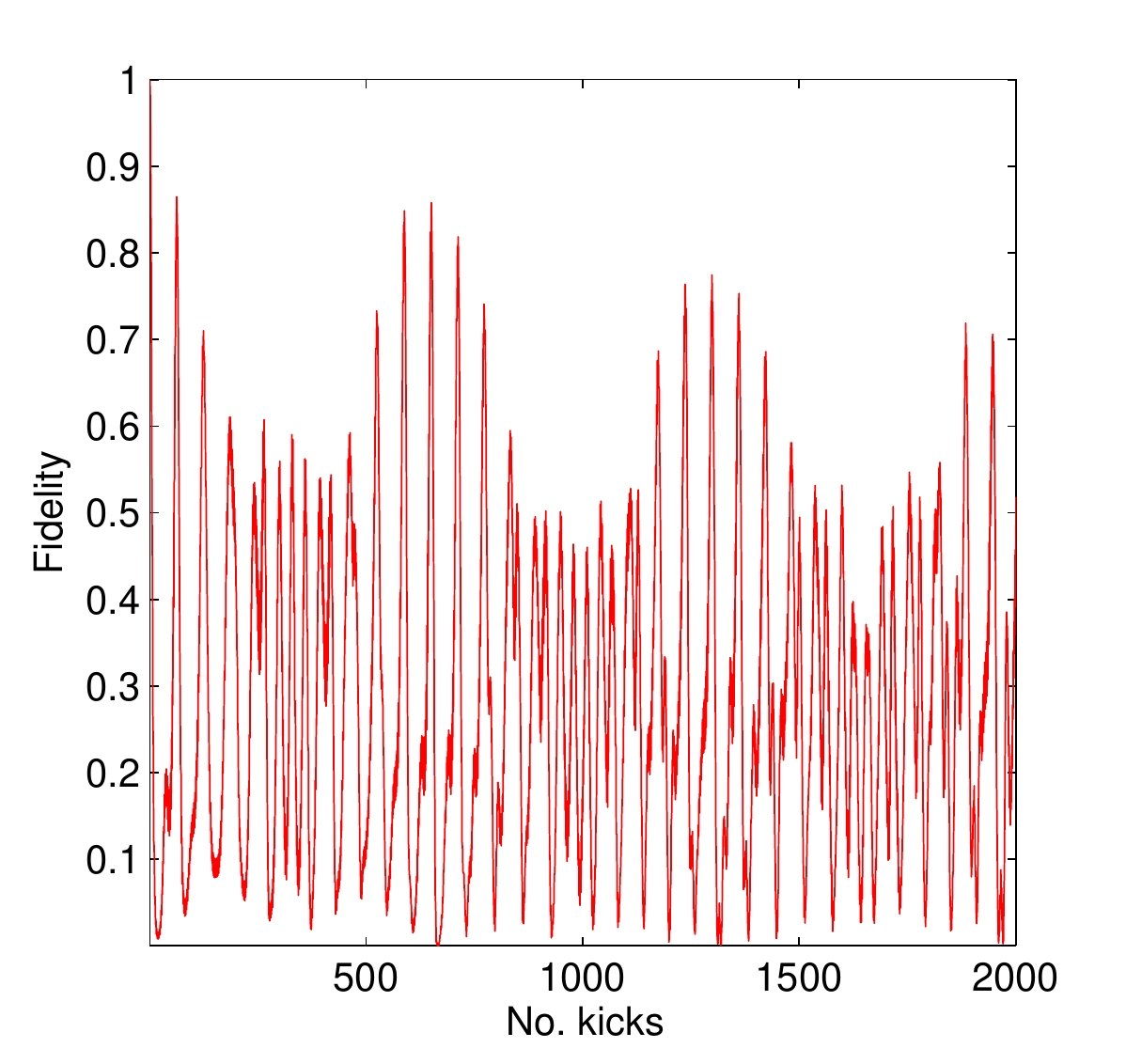}
\par\end{centering}

\caption{\label{fig:Fidelity-chain4}Fidelity of packet started inside an island
chain of order $4$. $K_{1}=2.10$, $K_{2}=2.11$, $\tau=2\times10^{-4}$,
and the center of the packet is started at $x=0.3198$, $p=0$, in
the center of one of the islands of the chain.}

\end{figure}

\begin{figure}
\begin{centering}
\includegraphics[width=8cm]{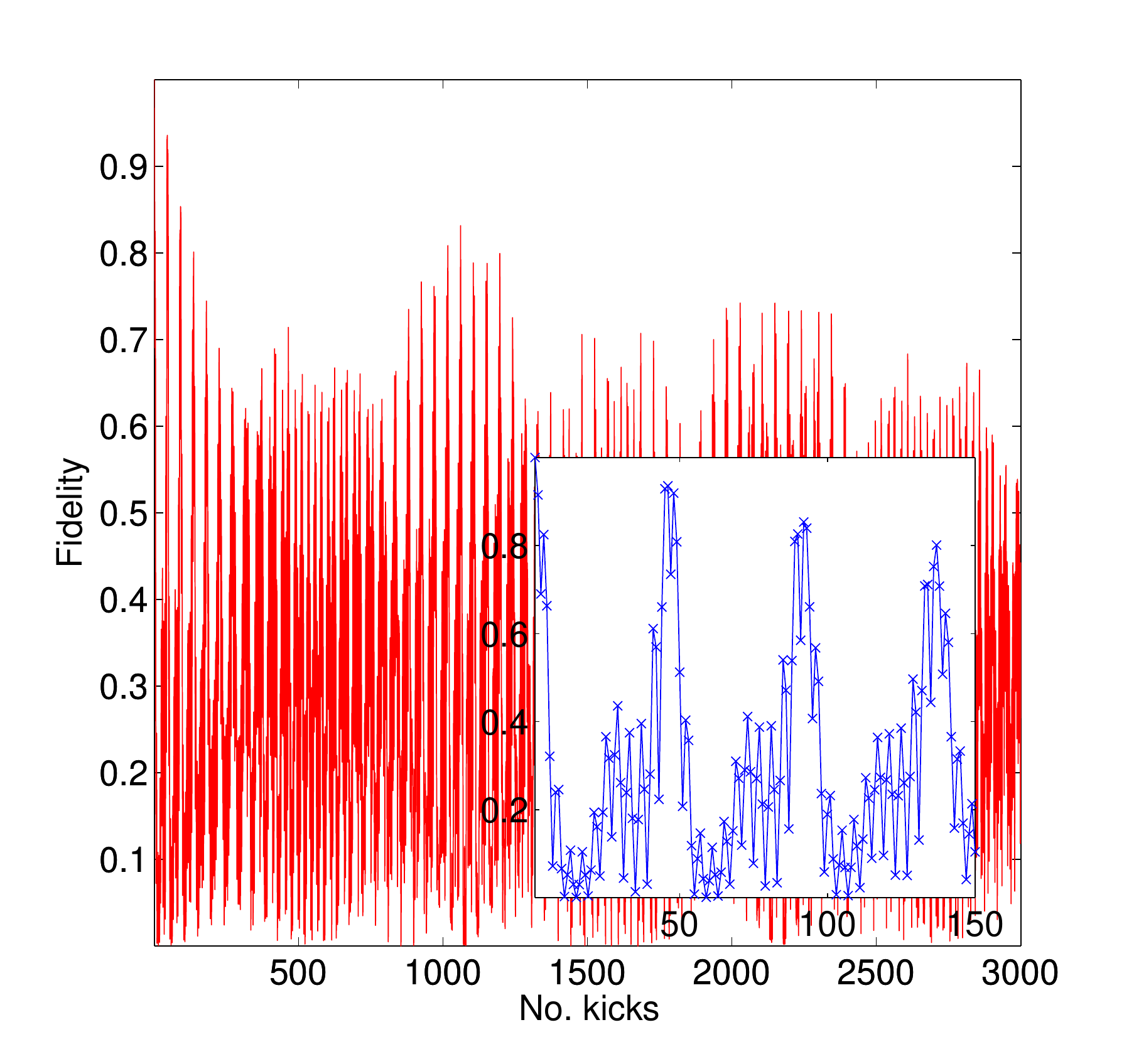}
\par\end{centering}

\caption{\label{fig:Fidelity-chain8}Fidelity of packet started inside an island
chain of order $8$. $K_{1}=1$, $K_{2}=1.01$, $\tau=2\times10^{-4}$,
and the center of the packet is started at $x=1.1312$, $p=0$, in
the center of one of the islands of the chain. The inset is a zoom
on the graph.}

\end{figure}

We consider two different island chains occurring for different values
of $K$. For $K=2.1$ we have examined a chain of order $r=4$ (see
Fig. \ref{fig:phase_space_K2.1}) %
, and for $K=1$ we have studied a chain of order $r=8$ (see Fig.
\ref{fig:phase_space_K1}). The initial wavepacket was launched inside
one of the islands of the chain, and the we numerically computed the
fidelity. In Figs. \ref{fig:Fidelity-chain4} ($K_{1}=2.10$, $K_{2}=2.11$)
and \ref{fig:Fidelity-chain8} ($K_{1}=1.00$, $K_{2}=1.01$)%
 we show the results of these computations.

It is notable that there are three timescales in the graph of the
fidelity. The shortest timescale is visible only in the inset of Fig.
\ref{fig:Fidelity-chain8} and may be understood taking into account
the symmetry of the equations of motion, $x\rightarrow-x,$ $p\rightarrow-p$.
This symmetry implies that each island has a {}``twin'' which is
found by reflection through the origin, $x=p=0$. Therefore, the overlap
between the islands of $H_{1}$ and $H_{2}$ is a periodic function
with a period of $\nicefrac{r}{2}$, where $r$ is the number of islands
in the chain. Consequently, for the island chains used to obtain Fig.
\ref{fig:Fidelity-chain4} and \ref{fig:Fidelity-chain8}, the fidelity
has periods of $2$ and $4$, respectively, on its shortest timescale.
The intermediate timescale is due to a rotation of the wavepacket
around the elliptic points of the island where it is initially launched.
The central point in the island is a fixed point of $M^{r}$. In $r$
iterations, points in the island rotate with an angular velocity $\omega_{1}$
and $\omega_{2}$ for $H_{1}$ and $H_{2}$, respectively. The angular
velocities can be calculated numerically by linearization of the tangent
map of $M^{r}$ around the fixed point of the map $M^{r}$. We find
the fixed point by reducing $M$ to a product of involutions \eqref{eq:involutions},
which allows us to reduce the search for the fixed points to the line
$p=0$ in the phase space since any point on this line is a fixed
point of $J_{1}$ \cite{Greene1979,Hihinashvili2007}. For $K_{1}=1$
and $K_{2}=1.01$, the angular velocities are found to be $\omega_{1}=1.10$
and $\omega_{2}=1.147$. For $K_{1}=2.10$ and $K_{2}=2.11$, the
angular velocities are found to be $\omega_{1}=0.391$ and $\omega_{2}=0.429$.
Therefore, the time it takes for a packet to accomplish a full revolution
around the fixed points of $M^{r}$ is $2\pi r/\bar{\omega}$, where
$\bar{\omega}=\frac{1}{2}\left(\omega_{1}+\omega_{2}\right)\approx\omega_{1}\approx\omega_{2}$
(see Table \ref{tab:chains_periods}). The longest timescale of the
fidelity is the timescale when the difference between the angular
velocities is resolved $T=2\pi r/\delta\omega$. In Table \ref{tab:chains_periods}
we compare those periods deduced directly from Fig. \ref{fig:Fidelity-chain4}
and Fig. \ref{fig:Fidelity-chain8} and the periods calculated by
finding $\omega_{1,2}$ from the tangent map. We see that the agreement
is excellent.%
\begin{table}
\begin{centering}
\begin{tabular}{ccccc}
\toprule
 & \multicolumn{2}{c}{$r=4$} & \multicolumn{2}{c}{$r=8$}\tabularnewline
\midrule
\midrule
 & Fig.\ref{fig:Fidelity-chain4} & Tangent map & Fig.\ref{fig:Fidelity-chain8} & Tangent map\tabularnewline
\midrule
shortest period & 2 & 2 & 4 & 4\tabularnewline
\midrule
medium period & 62 & $61.3$ & 44 & $44.7$\tabularnewline
\midrule
longest period & 651 & $657.8$ & 1061 & $1077.4$\tabularnewline
\bottomrule
\end{tabular}
\par\end{centering}

\caption{\label{tab:chains_periods}This table compares two ways of calculating
the periods of revivals for the resonance chains. In one way we have
deduced them from the Figures \ref{fig:Fidelity-chain4},\ref{fig:Fidelity-chain8},
and in the other way we have calculated them using the tangent map.
This is done for two different resonances: $r=4$, for $K_{1}=2.1$,
$K_{2}=2.11$; and $r=8$ for $K_{1}=1$, $K_{2}=1.01$. For both
cases $\tau=2\times10^{-4}$.}

\end{table}

\subsection{\label{sub:Fidelity_chaos}Fidelity of the wavepacket in the chaotic
strip}

\begin{figure}
\begin{centering}
\includegraphics[width=8cm]{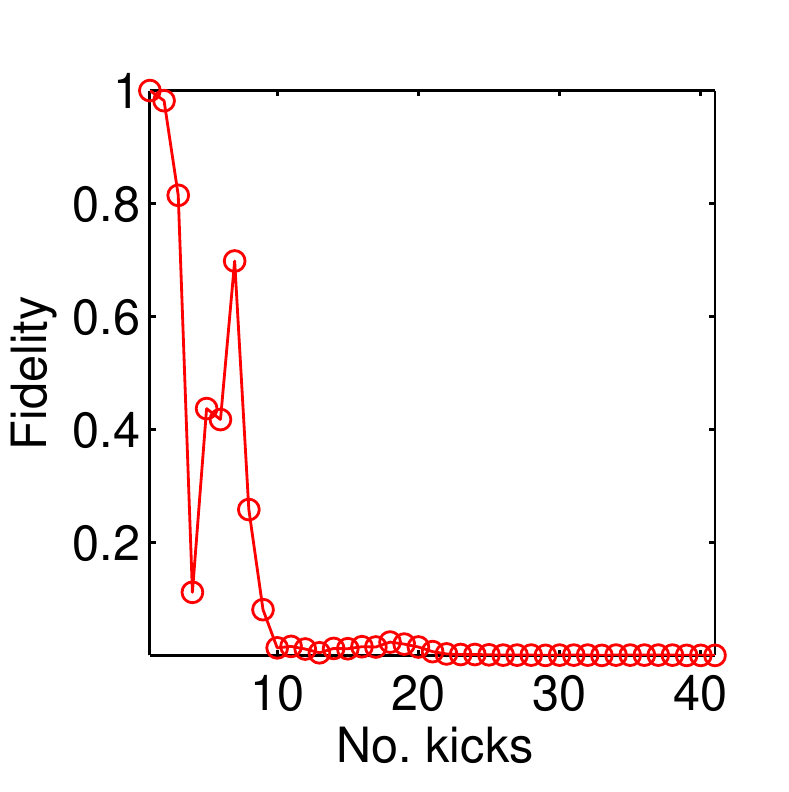}
\par\end{centering}

\caption{\label{fig:Fidelity-chaos}Fidelity of packet started inside a chaotic
layer. $K_{1}=1$, $K_{2}=1.01$, $\tau=2\times10^{-4}$ and the center
of the packet is started at $x=-2$, $p=0$.}

\end{figure}

For the fidelity of a packet started inside the chaotic strip (see
Fig. \ref{fig:Fidelity-chaos}), we notice a strong revival after
6 kicks which is dependent on $K_{1}$. This is half a period in this
chain/strip. After this revival the fidelity decays to zero, which
is a characteristic of chaotic regions.%
 Detailed exploration of this region is left for further studies.

\section{\label{sec:Dephasing}Dephasing}

\begin{figure}
\begin{centering}
\includegraphics[width=8cm]{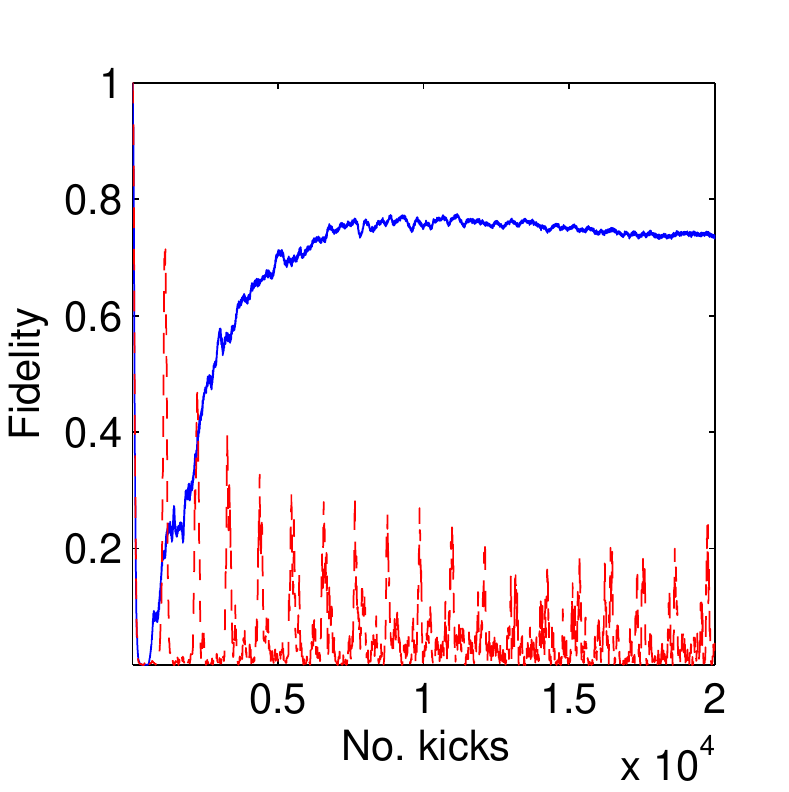}
\par\end{centering}

\caption{\label{fig:Quantum_classical_fidelity_noise0.01}(Color online) Quantum
fidelity, $S\left(t\right)$, (dashed red) and classical fidelity,
$S_{c}\left(t\right)$ (solid blue) for a dephasing noise of strength
$\sigma_{t}=0.01$, $K_{1}=1$, $K_{2}=1.01$, $\tau=0.01$, $x_{0}=-0.25$
and $p_{0}=0$.}

\end{figure}

\begin{figure}
\begin{centering}
\includegraphics[width=8cm]{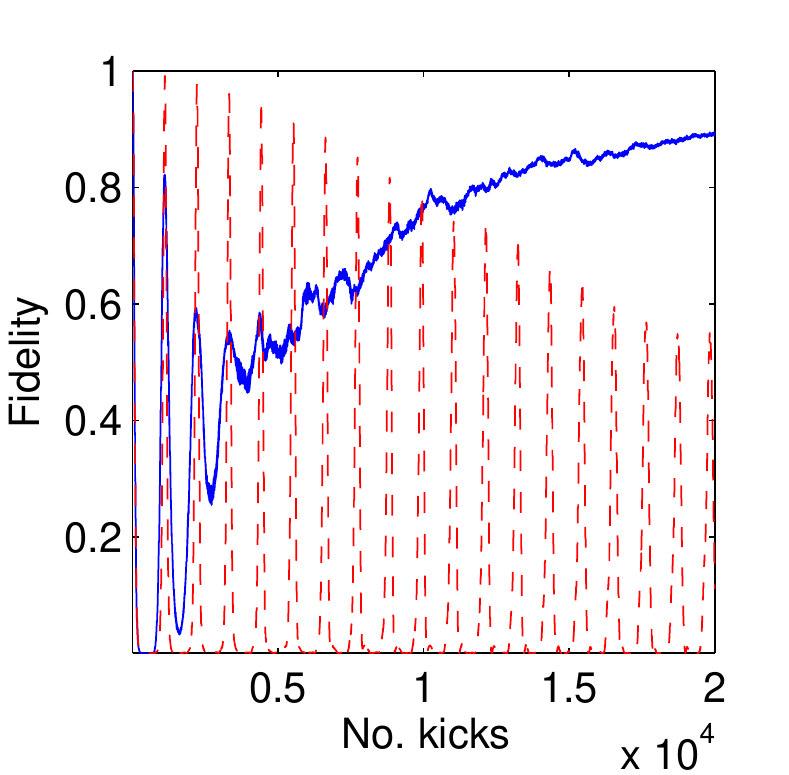}
\par\end{centering}

\caption{\label{fig:Quantum_classical_fidelity_noise0.001}(Color online) Quantum
fidelity, $S\left(t\right)$, (dashed red) and classical fidelity,
$S_{c}\left(t\right)$ (solid blue) for a dephasing noise of strength
$\sigma_{t}=0.001$, $K_{1}=1$, $K_{2}=1.01$, $\tau=0.01$, $x_{0}=-0.25$
and $p_{0}=0$.}

\end{figure}

We now investigate the effect of dephasing by adding temporal noise
to the time between the kicks. The classical equations of motion with
the dephasing are given by\begin{eqnarray}
p_{n+1} & = & p_{n}-Kx_{n}e^{-\frac{x_{n}^{2}}{2}},\\
x_{n+1} & = & x_{n}+\left(1+\delta t_{n}\right)\cdot p_{n+1},\nonumber \end{eqnarray}
and the quantum one kick propagator is\begin{equation}
U_{1}=e^{-i\frac{p^{2}}{2\tau}\left(1+\delta t_{n}\right)}\exp\left(i\frac{K}{\tau}e^{-\frac{x^{2}}{2}}\right),\end{equation}
where $\delta t_{n}$ is a random variable which is normally distributed
with zero mean and a standard deviation $\sigma_{t}$. The standard
deviation of the $\delta t_{n}$, corresponds to the strength of the
noise. We find that the noise results in an escape outside of the
island, which yields additional decay in the fidelity. Since we are
interested in the difference between the two wavefunctions only inside
the main island, for each kick we normalize the wavefunctions of $H_{1}$
and $H_{2}$ such that their norm is equal to $1$ inside a region
of $\left|x\right|\leq x_{b}=3$. This gives the following expression
for the fidelity\[
S\left(t\right)=\frac{\int_{-x_{b}}^{x_{b}}\left(e^{-iH_{1}t/\tau}\phi_{0}\left(x'\right)\right)\left(e^{-iH_{2}t/\tau}\phi_{0}\left(x'\right)\right)\, dx'}{\left(\int_{-x_{b}}^{x_{b}}\left|e^{-iH_{1}t/\tau}\phi_{0}\left(x'\right)\right|^{2}dx'\right)^{\nicefrac{1}{2}}\left(\int_{-x_{b}}^{x_{b}}\left|e^{-iH_{2}t/\tau}\phi_{0}\left(x'\right)\right|^{2}dx'\right)^{\nicefrac{1}{2}}},\]
with he classical fidelity $S_{c}\left(t\right)$ defined in a similar
way. We have numerically calculated the fidelity for the same situation
as in Fig. \ref{fig:Quantum_classical_fidelity} with added relative
noise of $\sigma_{t}=0.01$ (Fig. \ref{fig:Quantum_classical_fidelity_noise0.01})
and $\sigma_{t}=0.001$ (Fig. \ref{fig:Quantum_classical_fidelity_noise0.001}).
 We notice that the noise introduces additional decay in the quantum
fidelity.

\begin{figure}
\begin{centering}
\includegraphics[width=8cm]{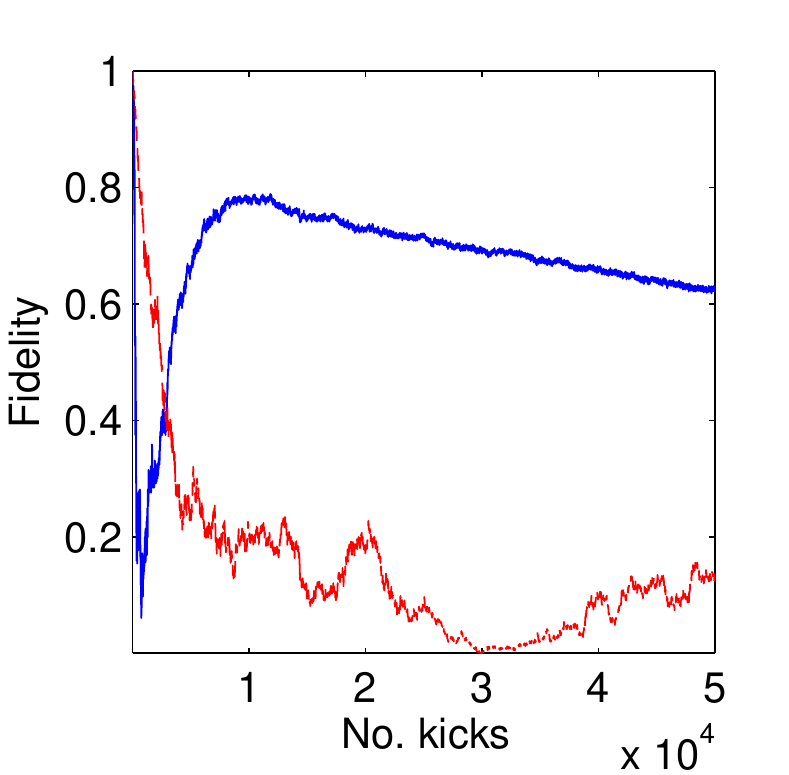}
\par\end{centering}

\caption{\label{fig:Quantum_classical_fidelity_noise0.01_sameK}(Color online)
Quantum fidelity, $S\left(t\right)$, (dashed light red) and classical
fidelity, $S_{c}\left(t\right)$ (solid dark blue) for two different
realizations of a dephasing noise of strength $\sigma_{t}=0.001$,
$K_{1}=K_{2}=1$, $\tau=0.01$, $x_{0}=-0.25$ and $p_{0}=0$.}

\end{figure}

\begin{figure}
\begin{centering}
\includegraphics[width=8cm]{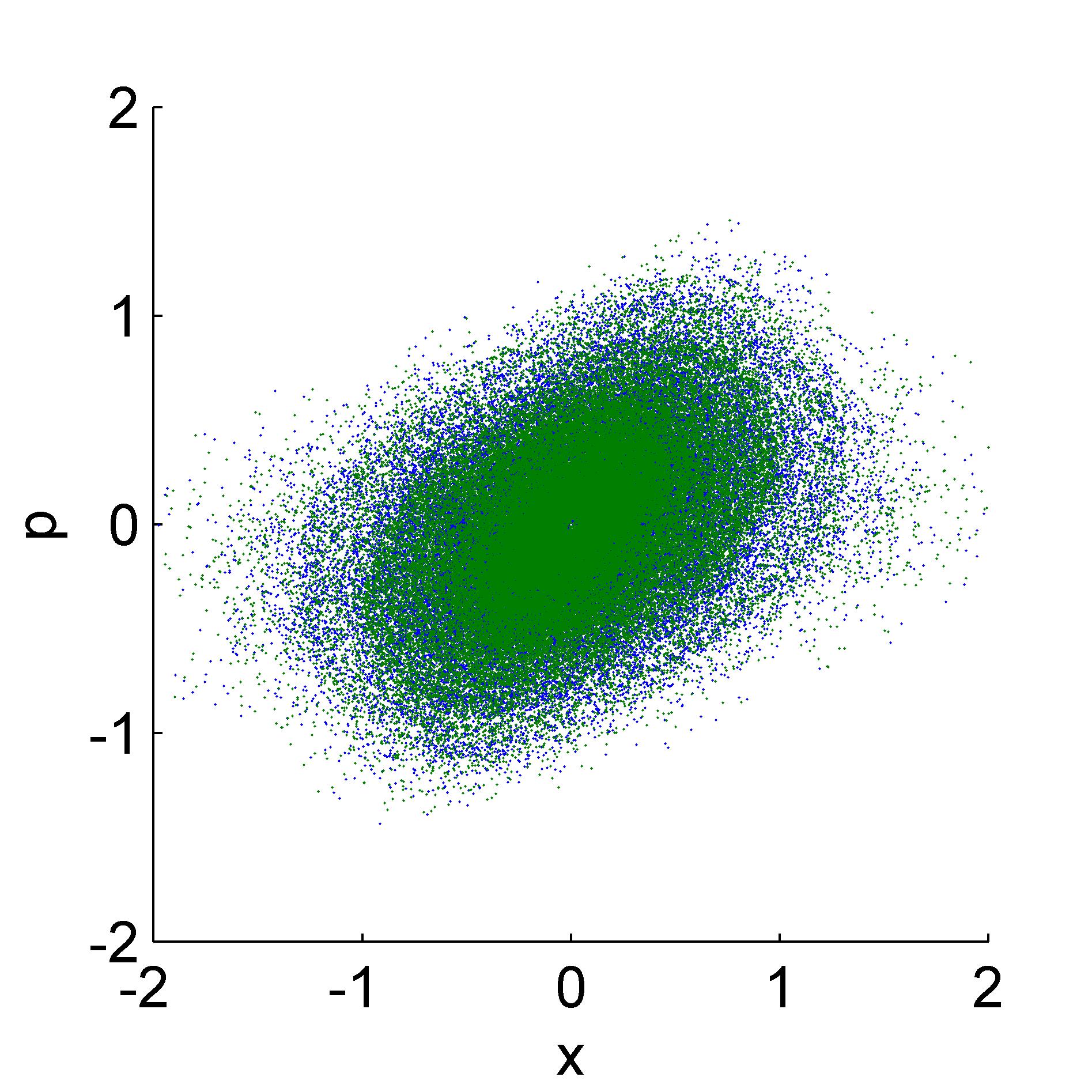}
\par\end{centering}

\caption{\label{fig:Whorls-sameK}(Color online) Classical density, which was
initially placed at $x_{0}=-0.25$ and $p_{0}=0$ after $5\times10^{4}$
kicks, $K_{1}=K_{2}=1$. Colors (shades) correspond to two different
realizations of a dephasing noise of strength $\sigma_{t}=0.01$.}

\end{figure}

\begin{figure}
\begin{centering}
\includegraphics[width=8cm]{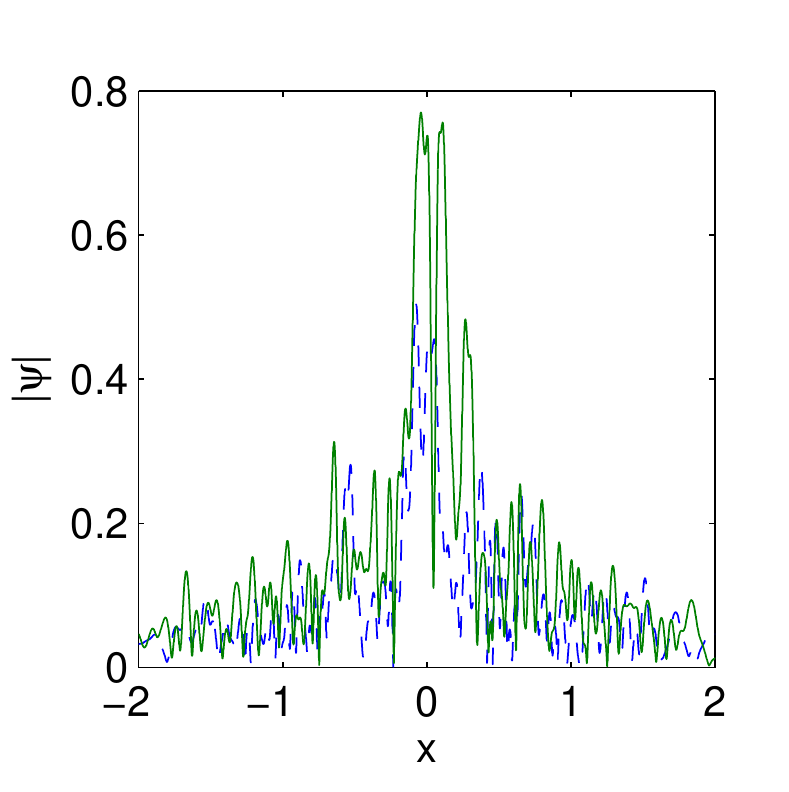}
\par\end{centering}

\caption{\label{fig:Psi-sameK}(Color online) Wavepackets, which were initially
placed at $x_{0}=-0.25$ and $p_{0}=0$ after $5\times10^{4}$ kicks,
$K_{1}=K_{2}=1$. Colors (shades) correspond to two different realizations
of a dephasing noise of strength $\sigma_{t}=0.01$.}

\end{figure}

To isolate the effect of noise from the decay in the fidelity due
to the difference between $K_{1}$ and $K_{2}$ we set $K_{1}=K_{2}$
and use two different noise realizations with the same strength $\sigma_{t}$.
From Fig. \ref{fig:Quantum_classical_fidelity_noise0.01_sameK} %
we notice that classical fidelity initially decays very fast due to
the noise and than slowly recovers approaching a value of $0.8$.
This is due to the coarse graining to the scale of $\tau$. To illustrate
this we plot in Fig. \ref{fig:Whorls-sameK} %
the classical densities after $5\times10^{4}$ kicks for a packet
initially launched at $x_{0}=-0.25$. We notice that the densities
for the two Hamiltonians highly overlap, which explains the high fidelity.
In Fig. \ref{fig:Psi-sameK} %
 we observe the corresponding quantum wavepackets. Contrary to the
classical fidelity, the quantum fidelity decays rather slowly with
the noise, suggesting that it is more robust to noise than the classical
fidelity.

\section{\label{sec:Scattering}Scattering}

We now investigate the difference between quantum and classical scattering
behavior by studying the evolution of a wavepacket initialized outside
of the main island of the phase space, Eq. \eqref{eq:Gaussian_quantum}
with $x_{0}=-2$, $p_{0}=0$, $\Delta x=\Delta p=\left(\tau/2\right)^{\nicefrac{1}{2}}$.
In the classical case both the classical chaos, as well as the numerous
small island structures introduce, an erratic behavior for the transmission
and reflection coefficients as a function of the initial launching
position and energy \cite{Jensen1992,Jensen1994}. Due to effective
phase space smoothing of areas much smaller than our effective Planck's
constant, $\tau$, we expect that fine scale fractal-like features
in the classically erratic scattering dependence will be averaged
out. To quantify this behavior, we measure the transmission and reflection
coefficients for a wavepacket defined as the transfered or reflected
probability mass, either quantum or classical. Classically, it is
the fraction of initial trajectories (generated using \eqref{eq:Gaussian_classical})
reflected or transmitted by the main island for a given time, while
quantum mechanically, we measure the total escaped probability up
to time $t$ from the island area, $\left|x\right|\leq x_{b}$, \begin{eqnarray}
L\left(t\right) & = & \int_{0}^{t}dt'\int_{-\infty}^{-x_{b}}\left|\psi\left(x,t'\right)\right|^{2}dx\\
R\left(t\right) & = & \int_{0}^{t}dt'\int_{x_{b}}^{\infty}\left|\psi\left(x,t'\right)\right|^{2}dx,\nonumber \end{eqnarray}
where $x_{b}$ is the margin of the main island (we choose $x_{b}=4$),
$L\left(t\right)$ and $R\left(t\right)$ are probabilities to be
scattered to the left or the right of the island till time $t$, correspondingly.
To determine those probabilities, we use the continuity equation for
the probability,\begin{equation}
\partial_{t}\left(\int_{a}^{b}\left|\psi\right|^{2}dx\right)=\tau\im\left[\left(\psi\partial_{x}\psi^{*}\right)|_{x=b}-\left(\psi\partial_{x}\psi^{*}\right)|_{x=a}\right],\end{equation}
so that,\begin{eqnarray}
L\left(t\right) & = & \frac{\tau}{2i}\int_{0}^{t}dt'\int_{0}^{t'}dt''\left(\psi\partial_{x}\psi^{*}-\psi^{*}\partial_{x}\psi\right)|_{x=-x_{b}},\\
R\left(t\right) & = & -\frac{\tau}{2i}\int_{0}^{t}dt'\left(\psi\partial_{x}\psi^{*}-\psi^{*}\partial_{x}\psi\right)|_{x=x_{b}}.\nonumber \end{eqnarray}

\begin{figure}
\begin{centering}
\includegraphics[width=8cm]{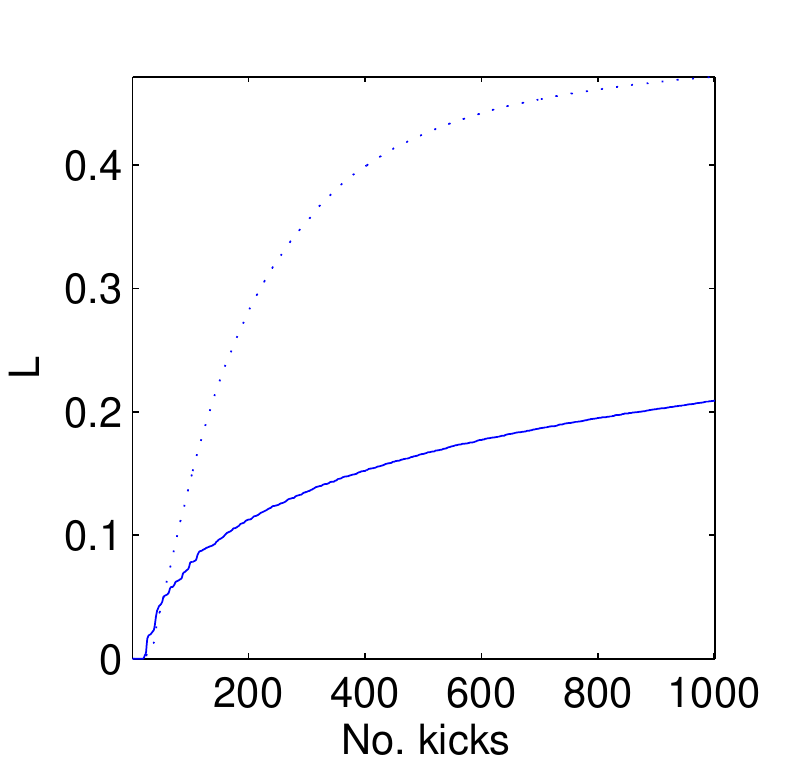}
\par\end{centering}

\caption{\label{fig:scatter_L}(Color online) Total quantum (solid blue line)
and classical (blue dots) probabilities for scattering to the left
of the island $\left(x<-x_{b}\right)$ as a function of the number
of kicks. $K=1$, $\tau=0.01$, $x_{0}=-2$, $p_{0}=0$.}

\end{figure}

\begin{figure}
\begin{centering}
\includegraphics[width=8cm]{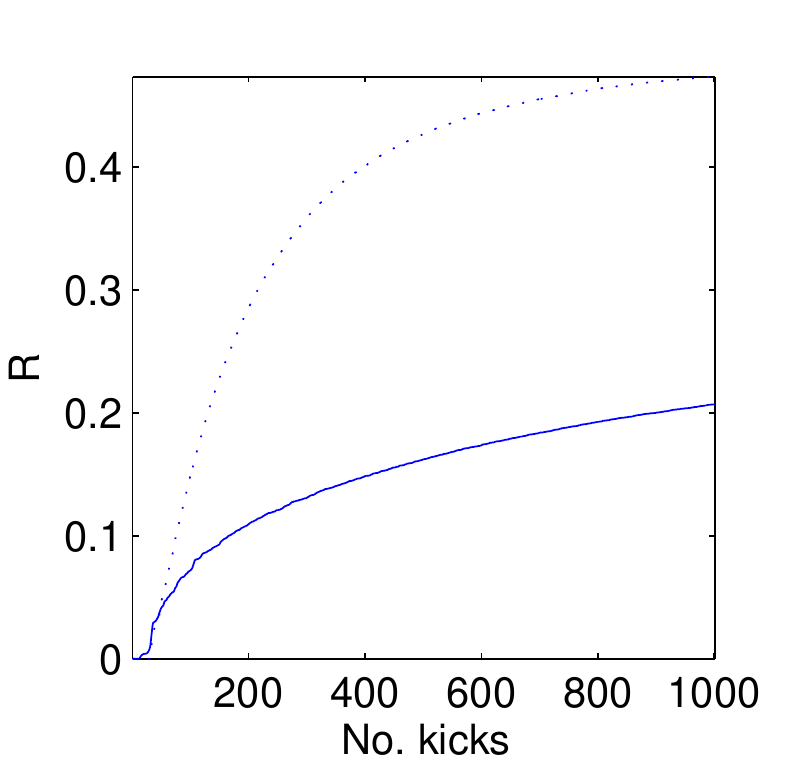}
\par\end{centering}

\caption{\label{fig:scatter_R}(Color online) Total quantum (solid blue line)
and classical (blue dots) probabilities for scattering to the right
of the island $\left(x>x_{b}\right)$ as a function of the number
of kicks. $K=1$, $\tau=0.01$, $x_{0}=-2$, $p_{0}=0$.}

\end{figure}

\begin{figure}
\begin{centering}
\includegraphics[width=8cm]{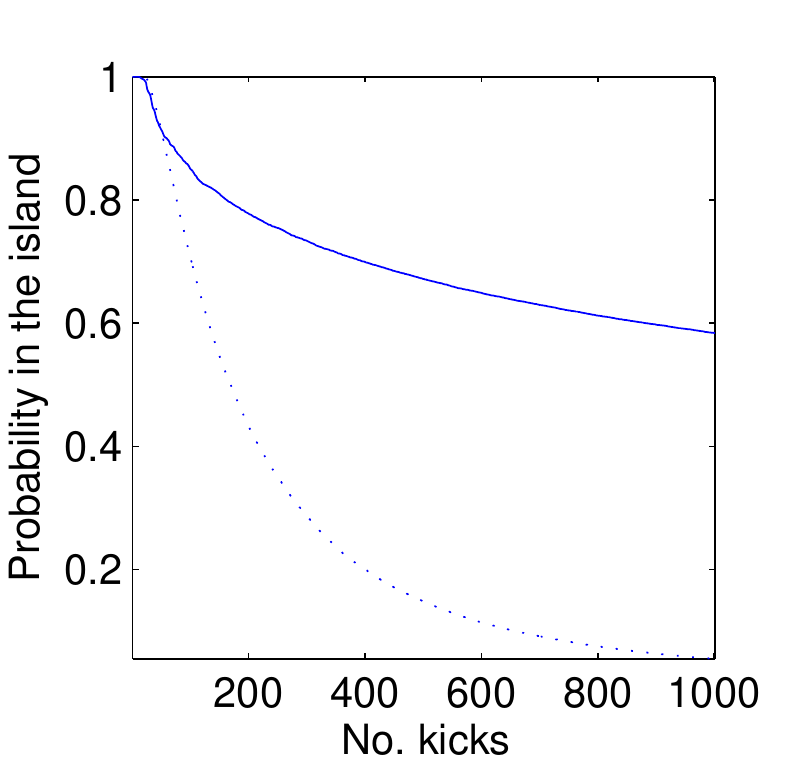}
\par\end{centering}

\caption{\label{fig:scatter_C}Total quantum (solid blue line) and classical
(blue dots) probabilities to stay in the island $\left(\left|x\right|\leq x_{b}\right)$
as a function of the number of kicks. $K=1$, $\tau=0.01$, $x_{0}=-2$,
$p_{0}=0$.}

\end{figure}

\begin{figure}
\begin{centering}
\includegraphics[width=8cm]{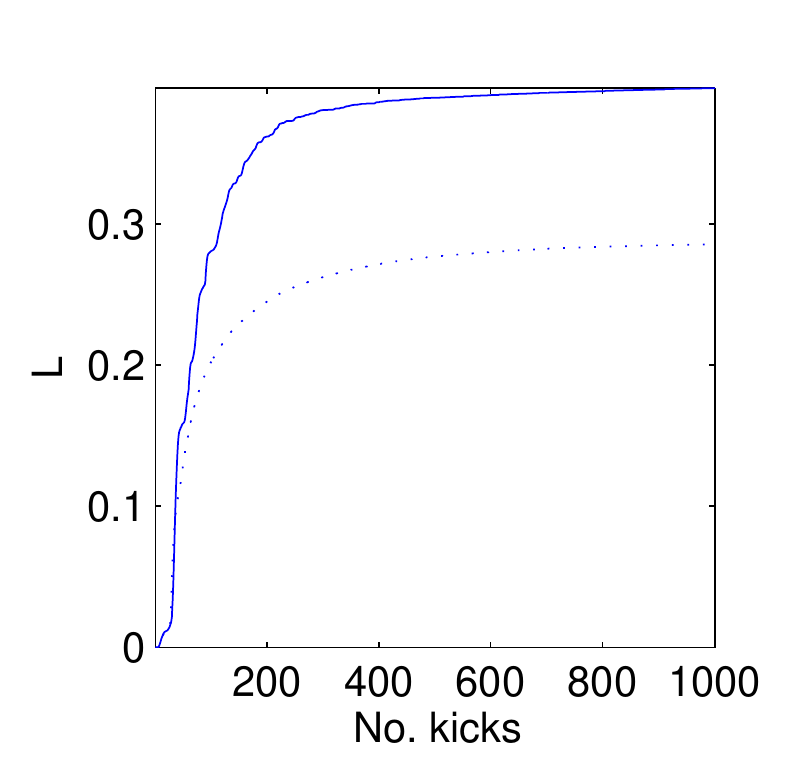}
\par\end{centering}

\caption{\label{fig:scatter_L-3}Total quantum (solid blue line) and classical
(blue dots) probabilities for scattering to the left of the island
$\left(x<-x_{b}\right)$ as a function of the number of kicks. $K=1$,
$\tau=0.01$, $x_{0}=-3$, $p_{0}=0$.}

\end{figure}

\begin{figure}
\begin{centering}
\includegraphics[width=8cm]{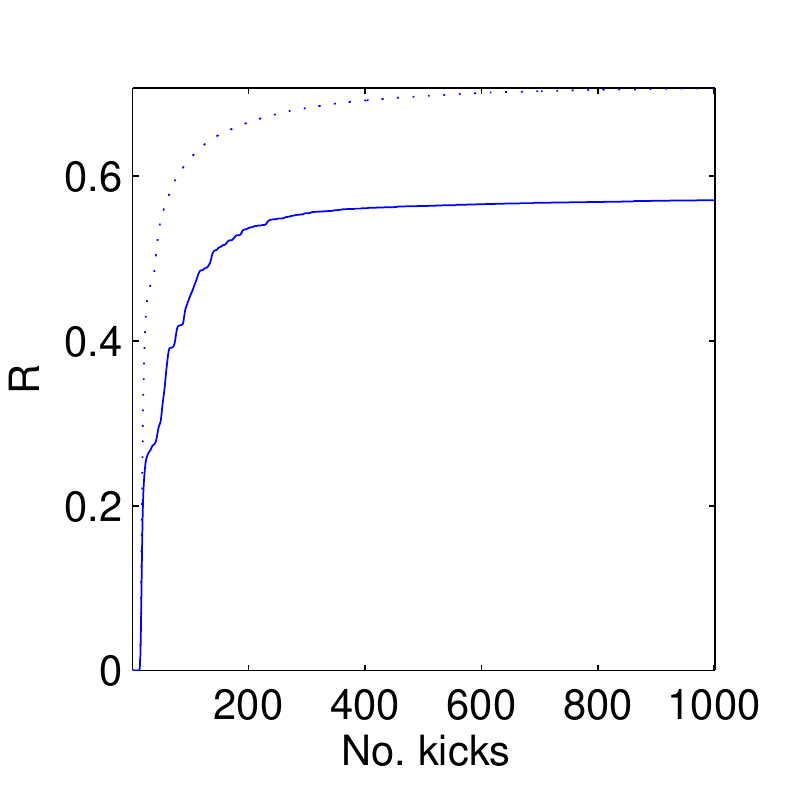}
\par\end{centering}

\caption{\label{fig:scatter_R-3}Total quantum (solid blue line) and classical
(blue dots) probabilities for scattering to the right of the island
$\left(x>x_{b}\right)$ as a function of the number of kicks. $K=1$,
$\tau=0.01$, $x_{0}=-3$, $p_{0}=0$.}

\end{figure}

\begin{figure}
\begin{centering}
\includegraphics[width=8cm]{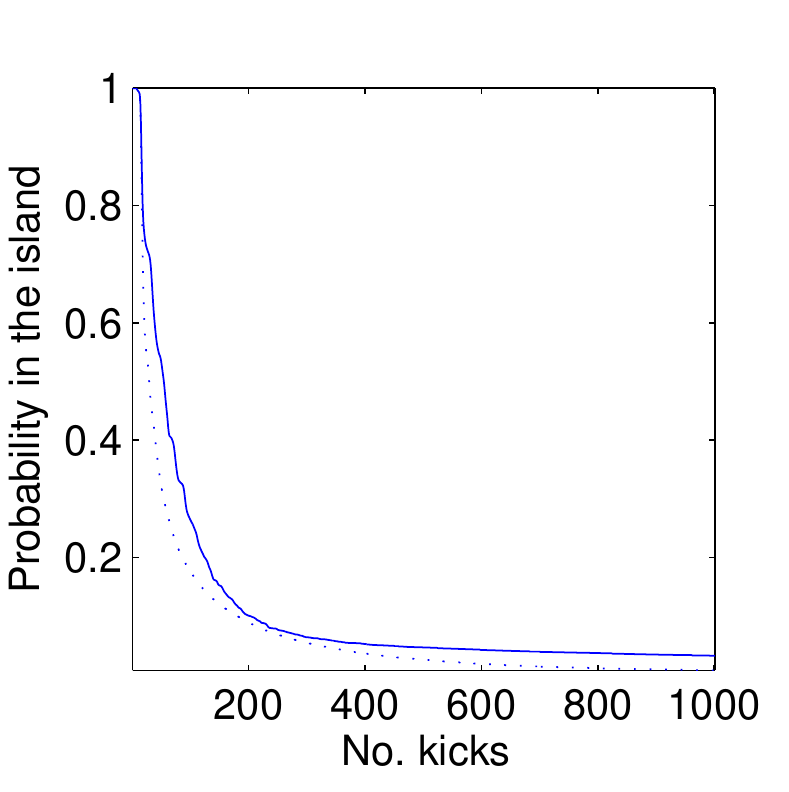}
\par\end{centering}

\caption{\label{fig:scatter_C-3}Total quantum (solid blue line) and classical
(blue dots) probabilities to stay in the island $\left(\left|x\right|\leq x_{b}\right)$
as a function of the number of kicks. $K=1$, $\tau=0.01$, $x_{0}=-3$,
$p_{0}=0$.}

\end{figure}

In Figs. \ref{fig:scatter_L}-\ref{fig:scatter_C-3}%
 we compare the quantum and classical scattering of a wavepacket launched
from the left of the main island. We notice that there is a substantial
difference, which decreases when we decrease the effective Planck's
constant, $\tau$. Figures \ref{fig:scatter_L}-\ref{fig:scatter_C}
and Figs. \ref{fig:scatter_L-3}-\ref{fig:scatter_C-3} differ in
the initial launching position of the wavepacket ($x_{0}=-2$, for
Figs. \ref{fig:scatter_L}-\ref{fig:scatter_C} and $x_{0}=-3$ for
Figs. \ref{fig:scatter_L-3}-\ref{fig:scatter_C-3}). We notice that
the scattering is sensitive to $x_{0}$. Different aspects of chaotic
scattering for this problem were explored in \cite{Jensen1992}, and
in particular, the effect of small $\hbar$ on washing out rainbow
singularities of the classical scattering function.

\section{\label{sec:Discussion}Discussion and Conclusions}

\subsection{Discussion of experimental realizability}

In the present work the classical and quantum dynamics of a system
with a mixed phase space were studied. It is proposed to realize this
system by injecting cold atoms into a coherent, pulsed, gaussian light
beam. The phase space structures, which can be seen on Figs. \ref{fig:phase_space_K1},\ref{fig:phase_space_K4.5}
and \ref{fig:phase_space_K2.1} are controlled by the parameters of
the beam via the parameter $K$. Since it is relatively straightforward
to control the parameters of gaussian beams, the proposed system is
ideal for the exploration of dynamics of mixed systems. In what follows
limitations on experimental realizations are discussed. First we consider
the realizability of an approximately one dimensional situation necessary
for the validity of our theoretical results. Let us assume that the
gaussian beam propagates in the $z$ direction. Its profile in the
$xy$ plane is \begin{equation}
e^{-\frac{x^{2}}{2\Delta^{2}}-\frac{y^{2}}{2\Delta_{y}^{2}}}.\label{eq:*1}\end{equation}
Assuming that the extent of the light beam is much smaller than the
Rayleigh length, $z_{R}=\pi\Delta^{2}/\lambda$ where $\lambda$ is
the wavelength, and the $z$ dependence of the potential can be ignored.
The potential of Eq. \eqref{eq:*1} can be well approximated by $\exp\left(-x^{2}/\left(2\Delta^{2}\right)\right)$
in \eqref{eq:Hamiltonian_with_units}, for sufficiently small values
of $y^{2}/\Delta_{y}^{2}$, and, to facilitate this, it is appropriate
to consider $\Delta_{y}\gg\Delta$, i.e., a quasi-sheet-like beam.
Such beams are experimentally realizable via routine methods. To analyze
this situation, the normalized map $M$ of \eqref{eq:classical_map}
should be replaced by one with $\exp\left(-x^{2}/2\right)$ replaced
by $\exp\left(-\left[x^{2}/2+y^{2}\left(\Delta/\Delta_{y}\right)^{2}\right]\right)$.
In addition, there are equations for $y_{n}$ and its conjugate momentum
$p_{y,n}$, which in dimensionless units with $y$ and $p_{y,n}$
rescaled by $\Delta$ and $T/\left(m\Delta\right)$, respectively,
take the form,\begin{eqnarray}
p_{y,n+1} & = & p_{y,n}-K_{y}y_{n}e^{-\frac{x_{n}^{2}}{2}-\frac{1}{2}\left(\frac{\Delta}{\Delta_{y}}\right)^{2}y_{n}^{2}},\nonumber \\
y_{n+1} & = & y_{n}+p_{y,n+1},\label{eq:*2}\end{eqnarray}
where

\begin{equation}
K_{y}=K\left(\frac{\Delta}{\Delta_{y}}\right)^{2}.\label{eq:*3}\end{equation}
Since $K\approx1$ and $\Delta/\Delta_{y}\ll1$, it can be assumed
that $K_{y}\ll1$. Therefore, the motion in the $y$ direction is
slow relative to the motion in the $x$ direction. Thus $\exp\left(-x_{n}^{2}/2\right)$
can be approximated by its time average $\left\langle \exp\left(-x^{2}/\left(2\Delta^{2}\right)\right)\right\rangle \equiv\rho$
(which is of order unity), and, for sufficiently small $y$ the $y-$motion
\eqref{eq:*2} can be described by a Harmonic oscillator with a force
constant $2K_{y}\rho\ll1$. Conservation of energy $E_{y}$ implies
that the maximal value of $y$ satisfies\begin{equation}
E_{y}=2K_{y}\rho y_{\mathrm{max}}^{2}.\label{eq:*5}\end{equation}
The energy $E_{y}$ is determined by the initial preparation. Let
us assume that initially the atoms form a Bose-Einstein Condensate
(BEC) and are in a harmonic trap that is anisotropic where the frequency
in the $y$ direction is $\nu_{y}^{\prime}$ in experimental units,
and $\nu_{y}=T\nu_{y}^{\prime}$ in our rescaled units. We assume
that the center of this trap $y_{0}$ satisfies $y_{0}\ll y_{max}$.
The experiment starts when the trap is turned off. Assuming the atoms
are in the ground state, their energy in our rescales units is\begin{equation}
\frac{1}{2}\hbar\nu_{y}^{\prime}\left(\frac{T^{2}}{m\Delta^{2}}\right)=\frac{1}{2}\nu_{y}\tau\leq E_{y}.\label{eq:*6}\end{equation}
We desire the effect of the motion in the $y$ direction on the motion
in the $x$ direction (Eq.\eqref{eq:Hamiltonian_with_units} with
$\exp\left(-x^{2}/\left(2\Delta^{2}\right)\right)$ replaced by $V$
of \eqref{eq:*1}) to be negligible. Thus it is required that\begin{equation}
\eta\equiv y_{\mathrm{max}}^{2}\left(\frac{\Delta}{\Delta_{y}}\right)^{2}\ll1.\label{eq:*7}\end{equation}
In this case the $y$ motion corresponds to a variation in $K$ of
the order $\Delta K\sim K\eta$. Using \eqref{eq:*5} and \eqref{eq:*6},
condition \eqref{eq:*7} reduces to\begin{equation}
\frac{1}{4}\frac{\nu_{y}\tau}{K}\leq\frac{E_{y}}{2K}=\eta\ll1,\label{eq:*8}\end{equation}
where, since we are interested only in crude estimates, we have replaced
$\rho$ by one. The initial spread in $y$ is given by the ground
state of the harmonic oscillator, where $\left\langle y^{2}\right\rangle =\tau/\left(2\nu_{y}\right)$,
and we require that the expectation value of $y^{2}$ satisfies $\left\langle y^{2}\right\rangle \ll y_{\mathrm{max}}^{2}$,
resulting in \begin{equation}
\left(\frac{\Delta}{\Delta_{y}}\right)^{2}\frac{\tau}{2\nu_{y}}\ll\eta.\label{eq:*9}\end{equation}
For both inequalities \eqref{eq:*8} and \eqref{eq:*9} to be satisfied
it is required that\begin{equation}
\left(\frac{\Delta}{\Delta_{y}}\right)^{2}\frac{\tau}{2\eta}\ll\nu_{y}\leq\frac{4K}{\tau}\eta.\label{eq:*10}\end{equation}
The resulting fundamental lower bound on $\eta$ is \begin{equation}
\left(\frac{\Delta}{\Delta_{y}}\right)^{2}\frac{\tau^{2}}{8K}\ll\eta.\label{eq:*11}\end{equation}
Reasonable experimental values are $\Delta/\Delta_{y}\approx10^{-2}$
and $\nu_{y}\approx0.1$. For $\tau=10^{-2}$ and $K\approx1$ the
lower bound on $\eta$ is $10^{-5}$ leaving a wide range for `engineering'
of BEC traps so that the $\nu_{y}$ is in the range \eqref{eq:*10}.
For $\nu_{y}\approx0.1$ and $\tau=10^{-2}$ and $K\approx1$ we can
make $\eta\lesssim10^{-3}$. Since this value of $\eta$ is small
compared to the value of $\Delta K=K_{2}-K_{1}$, used in our fidelity
calculations (Figs \ref{fig:Quantum_classical_fidelity}, \ref{fig:Fidelity_center},
\ref{fig:Quantum_classical_fidelity_noise0.01}, \ref{fig:Quantum_classical_fidelity_noise0.001}),
those calculations are expected to be uneffected by $y$ motion for
our assumed parameters. It is also encouraging to see that noise of
a higher level does not destroy fidelity oscillations (see Fig. \ref{fig:Quantum_classical_fidelity_noise0.001}).
One should note, however, that the variation of the effective $K$
of the motion in the $x$ direction is slow, with effective frequency
$\Delta/\Delta_{y}$ that for $\Delta/\Delta_{y}\approx10^{-2}$ is
of order $10^{-2}$. For these reasons, we expect that, the model
that we have explored theoretically in the present work should be
realizable for a wide range of experimental parameters.

\subsection{Conclusions}

The main result of this paper is that the quantum fidelity is sensitive
to the phase space details that are finer than Planck's constant,
contrary to expectations of Refs. \cite{Berry1972,Berry1977c}. In
particular, the fidelity was studied and predicted to oscillate with
frequencies that can be predicted from classical considerations. This
behavior is characteristic of regular regions. Fidelity exhibits a
periodic sequence of peaks. For wavepackets in the main island, it
was checked that the peak structure is stable in the presence of external
noise but the amplitude decays with time. For wavepackets initialized
in a chain of regular islands, it was found that the fidelity exhibits
several time scales that can be predicted from classical considerations.
For wavepackets initialized in the chaotic region, the fidelity is
found to decay exponentially as expected. It was shown how quasi-energies
are related to classical structures in phase space. Substantial deviation
between quantum and classical scattering was found. These quantum
mechanical effects can be measured with kicked gaussian beams as demonstrated
in the present work.
\begin{acknowledgments}
We are grateful to Steve Rolston for extremely detailed, informative
and critical discussions and Nir Davidson for illuminating comments.
This work was partly supported by the Israel Science Foundation (ISF),
by the US-Israel Binational Science Foundation (BSF), by the Minerva
Center of Nonlinear Physics of Complex Systems, by the Shlomo Kaplansky
academic chair, by the Fund for promotion of research at the Technion
and by the E. and J. Bishop research fund.
\end{acknowledgments}

\end{document}